\documentclass[aps,prd,showpacs,nofootinbib]{revtex4}
\usepackage{latexsym}
\usepackage{amsmath,amsfonts}
\usepackage{amsbsy}
\usepackage{mathrsfs}
\usepackage{color}

\usepackage{enumerate}

\usepackage{amsmath,amssymb,calc,amsfonts}
\usepackage{latexsym}

\usepackage{graphicx,calc,epsfig}
\def\ut#1{\rlap{\lower1ex\hbox{$\sim$}}#1{}}

\newcommand{\C}{\mathbb{C}}
\newcommand{\R}{\mathbb{R}}
\newcommand{\be}{\nopagebreak[3]\begin{equation}}
\newcommand{\ee}{\end{equation}}
\newcommand{\ba}{\nopagebreak[3]\begin{eqnarray}}
\newcommand{\ea}{\end{eqnarray}}
\DeclareFontFamily{U}{rsfs}{}         % Formal Script            %
\DeclareFontShape{U}{rsfs}{m}{n}{<5> rsfs5 <6><7> rsfs7          %
  <8><9><10><10.95><12><14.4><17.28><20.74><24.88> rsfs10}{}     %
\DeclareMathAlphabet{\mathfs}{U}{rsfs}{m}{n}                     %
\newcommand{\mfs}[1]{\mathfs {#1}}                               %
\newcommand{\inter}{{\lrcorner}}
\newcommand{\va}{\scriptscriptstyle}
\newcommand{\van}{\scriptstyle}

\newcommand{\sH}{{\mfs H}}
\newcommand{\sL}{{\mfs L}}

\newcommand{\sN}{{\mfs N}}
\newcommand{\sM}{{\mfs M}}

\newcommand{\sI}{{\mfs I}}
\newcommand{\n}{\nonumber}

\newcommand{\Lie}{\sL}

\newcommand{\Vect}{\mathrm{Vect}}

\newcommand{\eh}{|_{\va \Delta}=}

\def\pb#1{\rlap{\lower1.5ex\hbox{$\longleftarrow$}}{#1}}
\def\dpb#1{\rlap{\lower1.5ex\hbox{$\Longleftarrow$}}{#1}}
\def\spb#1{\rlap{\lower1.5ex\hbox{$\leftarrow$}}{#1}}
\def\sdpb#1{\rlap{\lower1.5ex\hbox{$\Leftarrow$}}{#1}}

\newcommand{\dummy}{\rule[0in]{0in}{0in}}

\begin{document}

\title{Black hole entropy from the $SU(2)$-invariant formulation of Type I isolated horizons}

\date{\today}

\author{Jonathan Engle $^{1,2}$}

\author{Karim Noui $^3$}

\author{Alejandro Perez$^1$}

\author{Daniele Pranzetti$^1$}

\affiliation{$^1$Centre de Physique Th\'eorique\footnote{Unit\'e
Mixte de Recherche (UMR 6207) du CNRS et des Universit\'es
Aix-Marseille I, Aix-Marseille II, et du Sud Toulon-Var; laboratoire
afili\'e \`a la FRUMAM (FR 2291)}, Campus de Luminy, 13288
Marseille, France.}

\affiliation{$^2$Institut f\"{u}r Theoretische Physik III,
Universit\"{a}t Erlangen-N\"{u}rnberg,
Staudtstra\ss e 7, 91058 Erlangen, Germany.}

\affiliation{$^3$ Laboratoire de Math\'ematiques et Physique
Th\'eorique\footnote{F\'ed\'eration Denis Poisson Orl\'eans-Tours,
CNRS/UMR 6083}, 37200 Tours, France.}

\begin{abstract}
A detailed analysis of the spherically symmetric isolated horizon system is performed in terms 
of the connection formulation of general relativity. The system is shown to admit a manifestly
$SU(2)$ invariant formulation where the (effective) horizon degrees of freedom are described 
by an $SU(2)$ Chern-Simons theory. This leads to a more transparent description of the quantum 
theory in the context of loop quantum gravity and modifications of the form of the horizon entropy. 
\end{abstract}

\pacs{04.60.-m, 04.60.Pp, 04.20.Fy, 11.15.Yc}

\maketitle

\section{Introduction}

Black holes are intriguing solutions of classical general relativity
describing important aspects of the physics of gravitational collapse.  Their existence in
our nearby universe is by now supported by a great amount of
observational evidence \cite{observ}.  When isolated, these systems are remarkably simple for late and distant observers: once
the initial very dynamical phase of collapse is passed the system is
expected to settle down to a stationary situation completely described (as implied by the famous results by Carter, Israel, and Hawking \cite{wald})
by the three extensive parameters (mass $M$, angular momentum $J$,
electric charge $Q$) of the Kerr-Newman family \cite{kerrnew}. 
 
However, the great simplicity of the final stage of an isolated
gravitational collapse for late and distant observers is in sharp
contrast with the very dynamical nature of the physics seen by
in-falling observers which depends on all the details of the
collapsing matter. Moreover, this dynamics cannot be consistently
described for late times (as measured by the infalling observers)
using general relativity due to the unavoidable development, within
the classical framework, of unphysical pathologies of the
gravitational field. Concretely, the celebrated singularity theorems
of Hawking and Penrose \cite{hawking} imply the breakdown of
predictability of general relativity in the black hole interior.
Dimensional arguments imply that quantum effects cannot be neglected
near the classical singularities. Understanding of physics in this 
extreme regime requires a quantum theory of gravity. Black holes (BH) provide,
in this precise sense, the most tantalizing theoretical evidence for
the need of a more fundamental (quantum) description of the
gravitational field.

Extra motivation for the quantum description of gravitational collapse
comes from the physics of black holes available to observers outside
the horizon.  As for the interior physics, the main piece of evidence
comes from the classical theory itself which implies an (at first only) apparent
relationship between the properties of idealized black hole systems and
those of thermodynamical systems.  On the one hand, black hole horizons satisfy
the very general Hawking area theorem (the so-called {\em second
law}) stating that the black hole horizon area $a_{\va H}$ can only
increase, namely \be \delta a_{\va H}\ge 0.  \ee On the other hand,
the uniqueness of the Kerr-Newman family, as the final (stationary)
stage of the gravitational collapse of an isolated gravitational
system, can be used to prove 
 the first and zeroth laws: under
external perturbation the initially stationary state of a black hole
can change but the final stationary state will be described by another
Kerr-Newman solution whose parameters readjust according to the {\em
first law} \be\delta M=\frac{\kappa_{\va H}}{8\pi G}\delta a_{\va
H}+\Phi_{\va H}\, \delta Q+\Omega_{\va H}\, \delta J,\ee where
$\kappa_{\va H}$ is the surface gravity, $\Phi_{\va H}$ is the
electrostatic potential at the horizon, and $\Omega_{\va H}$ the
angular velocity of the horizon.  There is also the {\em zeroth law}
 stating the
uniformity of the surface gravity $\kappa_{\va H}$ on the event
horizon of stationary black holes, and finally {\em the third law}
precluding the possibility of reaching an extremal black hole (for
which $\kappa_{\va H}=0$) by means of any physical
process\footnote{The third law can only be motivated by a series of
examples. Extra motivations comes from the validity of the cosmic
censorship conjecture.}.  The validity of these classical laws
motivated Bekenstein to put forward the idea that black holes may
behave as thermodynamical systems with an entropy $S=\alpha
a/\ell_p^2$ and a temperature $kT=\hbar \kappa_{\va H}/(8\pi \alpha)$
where $\alpha$ is a dimensionless constant and the dimensionality of
the quantities involved require the introduction of $\hbar$ leading in
turn to the appearance of the Planck length $\ell_p$, even though in
his first paper \cite{beke} Bekenstein states ``that one should not regard $T$ as the temperature of the black hole;
such identification can lead to all sorts of paradoxes, and is thus not useful''. The key point is that the
need of $\hbar$ required by the dimensional analysis involved in the
argument called for the investigation of black hole systems from a
quantum perspective.  In fact, not long after, the
semiclassical calculations of Hawking \cite{Hawking:1974sw}---that
studied particle creation in a quantum test field (representing
quantum matter and quantum gravitational perturbations) on the
space-time background of the gravitational collapse of an
isolated system described for late times by a stationary black hole---showed that once black holes
have settled to their stationary (classically) final states, they
continue to radiate as perfect black bodies at temperature
$kT=\kappa_{\va H}\hbar/(2\pi)$. Thus, on the one hand, this confirmed
that black holes are indeed thermal objects that radiate at a the
given temperature and whose entropy is given by $S = a/(4\ell^2_p)$,
while, on the other hand, this raised a wide range of new questions
whose proper answer requires a quantum treatment of the gravitational degrees
of freedom.

Among the simplest questions is the issue of the statistical origin of
black hole entropy. In other words, what is the nature of the the
large amount of micro-states responsible for black hole entropy. This
simple question cannot be addressed using semiclassical arguments of
the kind leading to Hawking radiation and requires a more fundamental
description. In this way, the computation of black hole entropy from
basic principles became an important test for any candidate quantum
theory of gravity.  In string theory it has been computed using
dualities and no-normalization theorems valid for extremal black holes
\cite{string}.  There are also calculations based on the effective
description of near horizon quantum degrees of freedom in terms of
effective $2$-dimensional conformal theories \cite{carlip}.  In loop
quantum gravity the first computations (valid for physical black
holes) were based on general considerations and the fact that the area
spectrum in the theory is discrete \cite{bhe0}.  The calculation was
later refined by quantizing a sector of the phase space of general
relativity containing a horizon in `equilibrium' with the external
matter and gravitational degrees of freedom \cite{bhe1}. In all cases
agreement with the Bekenstein-Hawking formula is obtained with
logarithmic corrections in $a/\ell^2_p$.

In this work we concentrate and further develop the theory of isolated horizons in the context  of loop quantum gravity.
Recently, we have proposed  a
new computation of BH entropy in loop quantum gravity
(LQG) that avoids the internal gauge-fixing used in
prior works \cite{nous} and makes the underlying structure more
transparent. We show, in particular, that the degrees of freedom of Type I 
isolated horizons can be encoded (along the lines of the standard treatment)
in an $SU(2)$ boundary connection.
The results of this work clarify the relationship
between the theory of isolated horizons and SU(2)
Chern-Simons theory first explored in \cite{kiril-lee}, and makes
the relationship with the usual treatment of degrees of freedom 
in loop quantum gravity clear-cut.
In the present work, we provide a full detailed derivation of 
the result of our recent work and discuss 
several important issues that  were only briefly mentioned then.

An important point should be emphasized concerning the logarithmic corrections mentioned above.
The logarithmic corrections to the Bekenstein-Hawking area formula for
black hole entropy in the loop quantum gravity literature were thought to be of the (universal)
form $\Delta S=-1/2 \log(a_H/\ell^2_p)$ \cite{amit}. In \cite{majundar} Kaul and Majumdar pointed out that, 
due to the necessary $SU(2)$ gauge symmetry of the isolated horizon system, the counting 
should be modified leading to corrections of the form $\Delta S=-3/2 \log(a_H/\ell^2_p)$.
This suggestion is particularly interesting because it would eliminate the apparent tension
with other approaches to entropy calculation. In particular their result is in complete agreement with the
seemingly very general treatment (which includes the string theory calculations) proposed by Carlip \cite{carlip-log}.
Our analysis confirms Kaul and Majumdar's proposal and eliminates in this way the apparent discrepancy 
between different approaches.

The article is organized as follows. In the following section we
review the formal definition of isolated horizons. In Section
\ref{mmain} we state the main equations implied by the isolated
horizon boundary conditions for fields at a spherically symmetric
isolated horizon. In Section \ref{CSS} we prove a series of
propositions that imply the main classical part of our results: we
derive the form of the conserved presymplectic structure of
spherically symmetric isolated horizons, and we show that degrees of
freedom at the horizon are described by an $SU(2)$ Chern-Simons
presymplectic structure.  In Section \ref{firstlaw} we briefly review
the derivation of the zeroth and first law of isolated horizons. In Section \ref{GSS} we study the gauge symmetries of the Type I isolated Horizon and explicitly compute the
constraint algebra. In Section \ref{quentum} we review
the quantization of the spherically symmetric isolated horizon phase
space and present the basic formulas necessary for the counting of
states that leads to the entropy.  We close with a discussion of our
results in Section \ref{conclu}. The appendix contains an analysis of
Type I isolated horizons from a concrete (and intuitive) perspective
that makes use of the properties of stationary spherically symmetric
black holes in general relativity.

\section{Definition of isolated horizons}\label{defini}

The standard definition of a BH as a spacetime region
from which no information can reach idealized observers
at (future null) infinity is a global definition. This notion
of BH requires a complete knowledge of a spacetime
geometry and is therefore not suitable for describing local
physics. The physically relevant definition used, for
instance, when one claims there is a black hole in the
center of the galaxy, must be local. One such local definition
was introduced in \cite{ack,better,ih_prl} with the name of isolated horizons (IH).
Here we present this definition according to 
\cite{ih_prl,afk,abl2002,abl2001}. This discussion will also serve to fix our notation.
In the definition of an isolated horizon below, we allow general matter, subject only to
conditions that we explicitly state.

\textit{Definition:} The internal boundary $\Delta$ of a history $(\sM, g_{ab})$ will be
called an \textit{isolated horizon} provided the following conditions hold:
\begin{enumerate}[i)]
\item \textit{Manifold conditions:} $\Delta$ is topologically $S^2 \times R$, foliated by a (preferred)
family of 2-spheres $S$ and equipped with an equivalence class $[\ell^a]$ of transversal
future pointing vector fields whose flow preserves the foliation, where $\ell^a$ is equivalent to 
$\ell'^a$ if $\ell^a = c \ell'^a$ for some positive real number $c$.

\item  \textit{Dynamical conditions:} All field equations hold at $\Delta$.

\item  \textit{Matter conditions:}
On $\Delta$ the stress-energy tensor $T_{ab}$ of matter is such that
$-T^a{}_b\ell^b$ is causal and future directed.

\item  \textit{Conditions on the metric $g$ determined by $e$, and on its levi-Civita derivative
operator $\nabla$:} (iv.a) The expansion of $\ell^a$ within $\Delta$ is zero.
%($\tilde{q}^{ab} \nabla_a \ell_b \equiv 2m^{(a} \bar{b}^{b)} \nabla_a \ell_b = 0$).
%
%  To do more justice to the robustness of this condition under different choices
%  inverse (equiv: different choices of $\ell^a$ and foliation), I would have to say much more.
%  But that is not appropriate.  This is not even explained in most of the original IH papers.
%  It also may be standard knowledge regarding congruences of null-geodesics.
%
%  Note, the above is the correct definition of expansion: see Robert Wald's GR book,
% and it is easy to see it is satisfied on IH knowing the properties you know about IH.
%
This, together with the energy condition (iii) and the Raychaudhuri equation at $\Delta$, ensures
that $\ell^a$ is additionally shear-free.  This in turn implies that the Levi-Civita derivative
operator $\nabla$ naturally determines a derivative operator $D_a$ intrinsic to $\Delta$ via
$X^a D_a Y^b := X^a \nabla_a Y^b$, $X^a, Y^a$ tangent to $\Delta$. We then impose (iv.b) $[\Lie_\ell, D] = 0$.
%
% Note: I had to put the matter conditions before the main conditions, so that I
% could continue to start with only non-expansion, and us the matter condition to
% ensure also $\ell^a$ is shear free, which ensures $D_a$ is well-defined, allowing
% us to state the second part of the main conditions.  This order is logical any way,
% because now all the very weak conditions are collected together as (i) - (iv), and
% the main non-trivial condition is condition (v).
%

\item  \textit{Restriction to `good cuts.'}
One can show furthermore that $D_a \ell^b = \omega_a \ell^b$ for some $\omega_a$ intrinsic to $\Delta$.
A 2-sphere cross-section $S$ of $\Delta$ is called a `good cut' if the pull-back of $\omega_a$ to $S$ is
divergence free with respect to the pull-back of $g_{ab}$ to $S$.  As shown in \cite{abl2002}, 
every horizon
satisfying (i)-(iv) above possesses at least one foliation into `good cuts'; this foliation is furthermore
generically unique.  We require that the fixed foliation coincide with a foliation into `good cuts.'

\end{enumerate}

Let us discuss the physical meaning of these conditions.
The first two conditions are rather weak.  The third condition is
satisfied by all matter fields normally used in general relativity.
The fifth condition is a partial gauge fixing of diffeomorphisms in the `time' direction.
The main condition is therefore the fourth condition.  
(iv.a) requires that $\ell^a$ be expansion-free.
This is equivalent to asking that the area 2-form of the 2-sphere cross-sections of $\Delta$
be constant along generators $[\ell^a]$. This combined with the matter condition (iii) and the Raychaudhuri equation
implies that in fact the \textit{entire} pull back $q_{ab}$ of the metric to the horizon is Lie dragged 
by $\ell^a$. Condition (iv.b) further stipulates that the derivative operator $D_a$ be Lie dragged by
$\ell^a$. This implies, among other things, an analogue of the \textit{zeroeth law} of black
hole mechanics: conditions (i) and (iii) imply that $\ell^a$ is geodesic ---
$\ell^b \nabla_b \ell_a \propto \ell_a$.  The proportionality constant is called the 
\textit{surface gravity}, and condition (iv.b) ensures that it is constant on the horizon for any 
given $\ell^a \in [\ell^a]$.  Furthermore, if we had not fixed $[\ell^a]$,
but only required that an $[\ell^a]$ exist such that the isolated horizon boundary
conditions hold, then condition (iv.b) would ensure that this $\ell^a$ is generically
unique \cite{abl2002}.
From the above discussion, one sees that the geometrical
structures on $\Delta$ that are time-independent are precisely the pull-back $q_{ab}$ of the 
metric to $\Delta$, and the derivative operator $D$. In fact, the main conditions (iv.a) and (iv.b) are equivalent
to requiring $\Lie_\ell q_{ab}=0$ and $[\Lie_\ell, D]=0$.
For this reason it is natural to define $(q_{ab}, D)$ as the \textit{horizon geometry}. 

%
% I decided not to get into mentioning any
% analogues of the laws of BH mechanics other than the zeroeth law,
% because none others are needed in this paper, and the others are
% not even mentioned in either of the entropy papers
% ACK and ABK, and what we are writing should be shorter!!!!!
%

Let us summarize.  Isolated horizons are null surfaces, foliated by a family
of marginally trapped 2-spheres such that certain geometric structures
intrinsic to $\Delta$ are time independent. The presence of trapped surfaces motivates
the term `horizon' while the fact that they are \textit{marginally} trapped ---
i.e., that the expansion of $\ell^a$ vanishes --- accounts for the adjective
`isolated'. The definition extracts from the definition of Killing horizon just that
`minimum' of conditions necessary for analogues of the laws of black hole mechanics to hold.
%
% This is stated, for example, in the entropy letter with Abhay and Chris Van Den Broeck.
%
Boundary conditions refer only to behavior of fields at $\Delta$ and
the general spirit is very similar to the way one formulates boundary conditions at null infinity.

\subsubsection*{Remarks:}
\begin{figure}[h]
\centerline{\hspace{0.5cm} \(
\begin{array}{c}
\includegraphics[height=4cm]{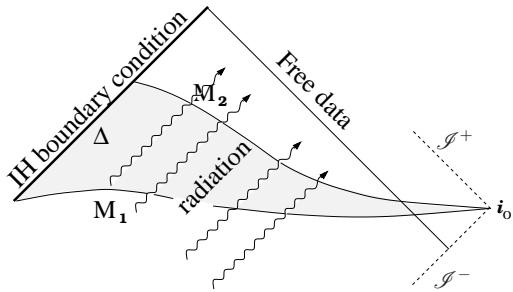}
\end{array}\!\!\!\!\!\!\!\!\!\!\!\! \!\!\!\!\!\!\!\!\!\!\!\! \begin{array}{c}  ^{}\\ \\  \\   \\   \\  \, \sI^{\va +}_{}  \\  \\ \\  \\  \\ \sI^{\va -}\end{array}\) } \caption{The characteristic data for a (vacuum) spherically symmetric
isolated horizon corresponds to Reissner-Nordstrom data on $\Delta$, and free radiation data
on the transversal null surface with suitable fall-off conditions. For each mass, charge, and radiation data in the transverse null surface there is a unique solution of
Einstein-Maxwell equations locally in a portion of the past domain of dependence of the null surfaces. 
This defines the phase space of Type I isolated horizons in Einstein-Maxwell theory. The picture shows two Cauchy surfaces $M_1$ and $M_2$ ``meeting'' at space-like infinity $i_0$.
A portion of $\sI^+$ and $\sI^-$ are shown; however, no reference to future time-like infinity $i^+$ is made as the isolated horizon need not to coincide with the black hole event horizon.}
\label{figui}
\end{figure}

\begin{enumerate}
\item
All the boundary conditions are satisfied by stationary black holes in the Einstein-Maxwell-dilaton
theory possibly with cosmological constant.
Note however that, in the non-stationary context, there still exist physically
interesting black holes satisfying our conditions: one can solve for all our conditions
and show that the resulting 4-metric need not be stationary on $\Delta$ \cite{lew2000}.

\item In the choice of boundary conditions, we have tried to strike the usual balance: On the one
hand the conditions are strong enough to enable one to prove interesting results (e.g.,
a well-defined action principle, a Hamiltonian framework, and a realization of black hole
mechanics) and, on the other hand, they are weak enough to allow a large class of examples.
As we already remarked, the standard black holes in the Einstein-Maxwell-dilatonic systems
satisfy these conditions.  More importantly, starting with the standard stationary black
holes, and using known existence theorems one can specify procedures to construct new
solutions to field equations which admit isolated horizons as well as radiation at null
infinity \cite{lew2000}. These examples, already show that, while the standard stationary
solutions have only a finite parameter freedom, the space of solutions admitting
isolated horizons is \textit{infinite} dimensional.  Thus, in the Hamiltonian picture,
even the reduced phase-space is infinite dimensional; the conditions thus admit a very
large class of examples.

\item Nevertheless, space-times admitting isolated horizon are 
very special among generic members of the full phase space of general relativity. The reason is
apparent in the context of the characteristic formulation of general
relativity where initial data are given on a set (pairs) of null
surfaces with non trivial domain of dependence. Let us take an
isolated horizon as one of the surfaces together with a transversal
null surface according to the diagram shown in Figure
\ref{figui}. Even when the data on the isolated horizon may be
infinite dimensional (for Type II and II isolated horizons, see
below), in all cases no transversing radiation data is allowed by the
IH boundary condition. Roughly speaking the isolated horizon boundary
condition reduces to one half the number of local degrees of freedom.

\item Notice that the above definition is completely geometrical and
does not make reference to the tetrad formulation. There is no
reference to any internal gauge symmetry. In what follows we will deal
with general relativity in the first order formulation which will
introduce, by the choice of variables, an internal gauge group
corresponding to local $SL(2,\C)$ transformations (in the case of
Ashtekar variables) or $SU(2)$ transformations (in the case of real
Ashtekar-Barbero variables).  It should be clear from the purely
geometric nature of the above definition that the IH boundary
condition cannot break by any means these internal symmetries.

\end{enumerate}
\subsection*{Isolated horizon classification according to their symmetry groups}

Next, let us examine symmetry groups of isolated horizons. A \textit{symmetry} of
$(\Delta, q, D, [\ell^a])$ is a diffeomorphism on $\Delta$ which preserves the horizon geometry
$(q, D)$ and at most rescales elements of $[\ell^a]$ by a positive constant. It is clear that 
diffeomorphisms generated by $\ell^a$ are symmetries. So, the symmetry group $G_\Delta$ is at 
least 1-dimensional. In fact, there are only three possibilities for $G_\Delta$:\\
\begin{enumerate}[(a)]
\item Type I: the isolated horizon geometry is spherical; in this case, $G_\Delta$ is four dimensional ($SO(3)$ rotations plus rescaling-translations\footnote{In a coordinate system where $\ell^a=(\partial/\partial v)^a$ the rescaling-translation corresponds to the affine map $v\to c v+b$ with $c,b \in \R$ constants. } along $\ell$);\\
\item Type II: the isolated horizon geometry is axi-symmetric; in this case, $G_\Delta$ is two dimensional
(rotations round symmetry axis plus rescaling-translations along $\ell$);\\
\item Type III: the diffeomorphisms generated by $\ell^a$ are the only symmetries; $G_\Delta$ is one dimensional.
\end{enumerate}

Note that these symmetries refer only to the horizon geometry. The
full space-time metric need not admit any isometries even in a
neighborhood of the horizon. In this paper, as in the classic works
\cite{bhe1, ack}, we restrict ourselves to the Type I
case. Although a revision would be necessary in light of the results of our present
work, the quantization and entropy calculation in the context of Type II and Type III
isolated horizons has been considered in \cite{jon}.

\section{Some extra details for Type I isolated horizons}\label{mmain}

In this section we first  list the main equations satisfied by fields
at an isolated horizon of Type I. The equations presented here can be directly derived
from the IH boundary conditions implied by the definition of Type I
isolated horizons given above. Most of the equations presented here  can be found in \cite{ack}. 
For completeness we prove these equations at the end of this section. { 
As we shall see in Subsection \ref{machin},  some of the coefficients entering the form of these equations
depend on the representative chosen among the equivalence class of null generators $[\ell]$. 
Throughout this paper we shall fix an null generator $\ell\in[\ell]$ by the requirement that the surface gravity
 $\ell \inter \omega=\kappa$ matches the one corresponding to the stationary black hole with the same macroscopic parameters as the 
 Type I isolated horizon under consideration. This choice makes the first law of IH take the form of the usual first 
 law of stationary black holes (see Section \ref{firstlaw}).}

\subsection{The main equations}

When written in connection variables, the isolated horizon boundary
condition implies the following relationship between the curvature of
the Ashtekar connection $A^i_{\va +}=\Gamma^i+i K^i$ at the horizon
and the $2$-form $\Sigma^i=\epsilon^{i}_{\ jk} e^j\wedge e^k$ (in the
time gauge) \be \sdpb{F_{ab}}^{i}(A^{\va +}) = -\frac{2\pi}{a_{\va
H}}\, \sdpb{\Sigma_{ab}}^{i}, \label{unos}\ee where $a_{\va H}$ is the
area of the IH, the double arrows denote the pull-back to
$H=\Delta\cap M$ with $M$ a Cauchy surface with normal
$\tau^a=(\ell^a+n^a)/\sqrt{2}$ at $H$, and $n^a$ null and normalized
according to $n\cdot\ell=-1$.  Notice that  the imaginary part of the previous equation implies that \be \sdpb{d_{\Gamma}K}^i=0\label{imi}\ee Another important equation is \be
\epsilon^{i}_{\ jk}\sdpb{K^j}\wedge \sdpb{K^k} =\frac{2\pi}{a_{\va H}}\sdpb{\Sigma}^i \label{doss}.
\ee The previous equations follow from  equations (3.12) and (B.7) of
reference \cite{ack}.  Nevertheless, they also follow from the abstract definition given in the introduction. From the previous equations, only equation (\ref{doss}) is not explicitly proven 
from the definition of IH in the literature. Therefore, we give here an explicit prove at the end of this section.
For concreteness, as we think it is helpful for some readers to have a concrete less abstract treatment, another derivation 
using directly the Schwarzschild geometry is given in
Appendix \ref{direct}.  The previous equations imply in turn that \be
\sdpb{F_{ab}}^{i}(A^{\va \beta}) = -\frac{\pi (1-\beta^2)}{a_{\va
H}}\, \sdpb{\Sigma_{ab}}^{i},\label{tress} \ee where $A^i_{\va
\beta}=\Gamma^i+\beta K^i$ is the Ashtekar-Barbero connection
\footnote{In our convention the $so(3)\to \R^3$ isomorphism is defined by $\lambda ^i=-\frac{1}{2}
\epsilon^{i}_{\ jk} \lambda^{jk}$ which implies that $F^i=dA^i+\frac{1}{2}
\epsilon^i_{\ jk} A^j\wedge A^k$ and
$d_A\lambda^i=d\lambda^i+\epsilon^i_{\ jk} A^j\wedge \lambda^k$.  }.

\subsection{Proof of the main equations}\label{machin}{
In this subsection we use the definition of isolated horizons provided in the previous section to prove
some of the equations stated above. { We will often work in a special gauge where the tetrad $(e^I)$ is such that $e^1$ is normal 
to $H$ and $e^2$ and $e^3$ are tangent to $H$. This choice is only made for convenience, as the equations presented in the previous section are all gauge
covariant, their validity in one frame implies their validity in all frames.}

 \noindent {\bf Lemma 1:}  In the gauge where the tetrad is chosen so that  $\ell^a=2^{-1/2} (e^a_0+e^a_1)$ (which can be completed to a null tetrad $n^a=2^{-1/2} (e^a_0-e^a_1)$, and $m^a=2^{-1/2} (e^a_2+i e^a_3)$),  the  shear-free and vanishing expansion (condition ($iv.a$) in the definition of IH)  imply
\be \sdpb{\omega}^{21}=\sdpb{\omega}^{20} \ \ \ {\rm and}\ \ \ \sdpb{\omega}^{31}=\sdpb{\omega}^{30}. \label{l1}\ee 
\noindent {\it Proof:}
The expansion $\rho$ and shear $\sigma$ of the null congruence of generators $\ell$ of the horizon is given by 
\be
\rho=m^a\bar m^b \nabla_a\ell_b, \ \ \ \ \ \ \sigma=m^a m^b \nabla_a\ell_b.
\ee 
This implies
\ba
0&=&\rho=\frac{1}{2\sqrt{2}} m^a(e^b_2-ie^b_3)\nabla_a (e_b^1-ie_b^0)=\\
&=&\frac{1}{2\sqrt{2}}m^a((\omega^{21}_a-\omega^{20}_a)-\imath(\omega^{31}_a-\omega^{30}_a)),
\ea
where we have used the definition of the spin connection $\omega_a^{IJ}=e^{I b}\nabla_a e_b^J$.
Similarly we have 
\ba
0&=&\sigma=\frac{1}{2\sqrt{2}} m^a(e^b_2+ie^b_3)\nabla_a (e_b^1-ie_b^0)=\\
&=&\frac{1}{2\sqrt{2}}m^a((\omega^{21}_a-\omega^{20}_a)+\imath(\omega^{31}_a-\omega^{30}_a)).
\ea
As $e_2^a$ and $e_3^a$ form a non degenerate frame for $H=\Delta\cap M$, and from the definition of pull-back,  the previous two equations imply the statement of our lemma. $\square$

The previous lemma has an immediate consequence on the form of equation (\ref{imi}) for the component $i=1$ in the frame of the previous lemma. More precisely it says that  $\sdpb{dK}^1=0$. The good-cut condition ($v$) in the definition implies then that 
\be
\sdpb{K}^1=0.\label{k1}
\ee
Another important consequence of the previous lemma is equation (\ref{unos}), also derived in \cite{ack}. We give here for completeness and self consistency a sketch  of its derivation.
This equation follows from identity
\be
F_{ab}{^i}(A^{\va +})=-\frac{1}{4} R_{ab}^{\ \ cd} \Sigma^{\va + i}_{cd},
\ee
where $R_{abcd}$ is the Riemann tensor and $\Sigma^{\va + i}= \epsilon^{i}_{\ jk} e^j\wedge e^k+ i2 e^0\wedge e^i$, which can be derived using Cartan's structure equations.
A simple algebraic calculation using the null tetrad formalism (see for instance \cite{chandra} page 43) with the 
null tetrad of Lemma 1, and the definitions $\Psi_2=C_{abcd} \ell^a m^b \bar m^c n^d$ and $\Phi_{11}=R_{ab}(\ell^an^b+m^a\bar{m}^b)/4$,
 where $R_{ab}$ is the Ricci tensor and $C_{abcd}$ the Weyl tensor, yields 
 \be
 \sdpb{F_{ab}}^i=(\Psi_2-\Phi_{11}-\frac{R}{24})\sdpb{\Sigma}^i,
 \ee
 where $\Sigma^{i}={\rm Re}[\Sigma^{{\va +}i}]= \epsilon^{i}_{\ jk} e^j\wedge e^k $.
 An important point here is that the previous expression is valid for any two sphere $S^2$ embedded in spacetime 
 in an adapted null tetrad where $\ell^a$ and $n^a$ are normal to $S^2$. However, in the special case where $S^2=H$
 (where $H=\Delta\cap M$ with $\Delta$ a Type I isolated horizon) it follows from spherical symmetry that $(\Psi_2-\Phi_{11}-\frac{R}{24})=C$ with $C$ a constant 
 on the horizon $H$.  Moreover, in the gauge defined in the statement of Lemma 1, the only non vanishing component of the 
 previous equation is the $i=1$ component for which (using Lemma 1) we get \be dA_{\va +}^1=C \epsilon, \ee with $\epsilon$ the area element of $H$.
Integrating the previous equation on $H$ one can completely determine the constant $C$, namely
 \be
 C=(\Psi_2-\Phi_{11}-\frac{R}{24})=-\frac{2\pi}{a_{\va H}},
 \ee
 from where equation (\ref{unos}) immediately follows.
 
 \noindent {\bf Lemma 2:} For Type I isolated horizons
\be\label{eq:KK=lambda_0 Sigma}
\sdpb{K^j}\wedge \sdpb{K^k}\epsilon_{ijk}=c_0 \frac{2\pi }{a_{\va H}} \sdpb{\Sigma}^i\, ,
\ee 
for some constant $c_0$. One can choose a representative from the equivalence class $[\ell]$ of null normals to the isolated horizon in order to 
fix $c_0=1$ by making use of the translation symmetry of IH along $\ell$. { By studying the stationary spherically symmetric back hole solutions one
can show that this corresponds to the choice where the surface gravity of the IH matches the stationary surface gravity (see Appendix \ref{direct}).}
\vskip.4cm
\noindent {\it Proof:} In order to simplify the notation all free indices associated to forms that appear in this proof are pulled back to $H$ (this allows us to drop the double arrows from equations).
In the frame of lemma 1,  where  $e^1$ is normal to $H$,  the only non trivial component of the  equation  we want to prove  is
the $i=1$ component, namely:
 \be
{K^A}\wedge {K^B}\epsilon_{AB}=c_0 \frac{2\pi }{a_{\va H}} {\Sigma}^1 , \label{l2}\ee 
 where $A,B=2,3$ and $\epsilon^{AB}=\epsilon^{1AB}$. Now,  in that gauge,  we have that $K^A=c^A_{\ B} e^B$ for some matrix of coefficients $c^A_{\ B}$. 
 Notice that the left hand side of the previous equation equals $\det(c) e^A\wedge e^B\epsilon_{AB}$.
We first prove that $\det(c)$ is time independent, i.e. that $\ell(\det{c})=0$.
%To prove this it's enough to show that  
%\be\label{eq:Boundary Condition K}
%\mathscr{L}_\ell K^i=0.
%\ee
We need to use the isolated horizon boundary condition 
\be
[\mathscr{L}_\ell, D_b] v^a=0 \ \ \ v^a\in T(\Delta)
\ee
where $D_a$ is the derivative operator determined on the horizon by the Levi-Civita derivative operator $\nabla_a$. 
One important property of the commutator of two derivative operators is that it also satisfy the Leibnitz rule (it is itself a new derivative operator). Therefore, using the fact that  the null vector
$n^a$ is normalized so that $\ell\cdot n=-1$ we get
\be
0=[\mathscr{L}_\ell, D_b] \ell^a n_a=n_a[\mathscr{L}_\ell, D_b] \ell^a +\ell^a [\mathscr{L}_\ell, D_b] n_a \ \ \ \Rightarrow \ \ \ \ell^a [\mathscr{L}_\ell, D_b] n_a=0,\label{12}
\ee
where we have also used that $\ell^a\in T(\Delta)$.
Evaluating the equation on the right hand side explicitly, and using the fact that $\mathscr{L}_\ell n=\ell \inter dn+d(\ell \inter n)=0$\footnote{ Here we used that $dn = 0$ which comes from  the restriction to `good cuts'
in definition of Section \ref{defini}. More precisely, if one introduces a coordinate
$v$ on $\Delta$ such that $\ell^a \partial_a v = 1$ and
$v = 0$ on some leaf of the foliation, then it follows---from the fact that 
$\ell$ is a symmetry of the horizon geometry $(q,D)$,
and the fact that the horizon geometry uniquely determines
the foliation into `good cuts'---that $v$ will be constant on
all the leaves of the foliation.  As $n$ must be normal to the leaves one has  $n = - dv$, whence
$dn = 0$.} we get
 \ba\n 0&=& \ell^a [\mathscr{L}_\ell, D_b] n_a =\ell^a \mathscr{L}_\ell (D_b n_a)=-\frac{1}{\sqrt{2}}\ell^a \mathscr{L}_\ell (D_b [e^1_a+e^0_a]) \\ 
 &=& \frac{1}{\sqrt{2}}\ell^a \mathscr{L}_\ell (\omega^1_{b\ \mu} e^{\mu}_a+\omega^{0}_{b\ \mu}e^{\mu}_a)= -\frac{1}{\sqrt{2}}\ell^a \mathscr{L}_\ell (\omega_b^{10} [e^{0}_a+e^{1}_a])+\frac{1}{\sqrt{2}}\ell^a \mathscr{L}_\ell(\omega^1_{b\ A} e^{A}_a+\omega^{0}_{b\ A}e^{A}_a)\n \\
 &=& \ell^a \mathscr{L}_\ell (\omega_b^{10}) n_a ,\n 
\ea where in the second line we have used the fact that $D_be_a^\nu=-\omega^{\nu}_{b\ \mu} e_a^{\mu}$ plus the fact that as $\sL_{\ell}q_{ab}=0$ the Lie derivative $\sL_{\ell} e^{A}=\alpha \epsilon^{AB} e_{B}$ for some $\alpha$ (moreover, one can even fix $\alpha=0$ if one wanted to by means of internal gauge transformations). Then it follows that
\be
\sL_{\ell} K^1=0,
\ee
a condition that is also valid for the so called {\em weakly isolated horizons} \cite{better}. A similar argument as the one given in equation (\ref{12})---but now replacing $\ell^a$ by $e^a_{B}\in T(\Delta)$ for $B=2,3$---leads to
\ba\n 0&=& e_B^a [\mathscr{L}_\ell, D_b] n_a =e_B^a \mathscr{L}_\ell (D_b n_a)=-\frac{1}{\sqrt{2}}e_B^a \mathscr{L}_\ell (D_b [e^1_a+e^0_a]) \\ 
 &=& \frac{1}{\sqrt{2}}e_B^a \mathscr{L}_\ell (\omega^1_{b\ \mu} e^{\mu}_a+\omega^{0}_{b\ \mu}e^{\mu}_a)= -\frac{1}{\sqrt{2}}e_B^a \mathscr{L}_\ell (\omega_b^{10} [e^{0}_a+e^{1}_a])+\frac{1}{\sqrt{2}}e_B^a \mathscr{L}_\ell(\omega^1_{b\ A} e^{A}_a+\omega^{0}_{b\ A}e^{A}_a)\n \\
 &=& {\sqrt{2}}e_B^a \mathscr{L}_\ell(\omega^{0}_{b\ A}e^{A}_a)={\sqrt{2}} [\mathscr{L}_\ell(\omega_b^{0B}) + \alpha \epsilon^{BA} \omega_b^{0A}]\n,
\ea where, in addition to previously used identities, we have made use of lemma 1, eq. (\ref{l1}). The previous equations imply that the left hand side of equation (\ref{l2}) is Lie dragged along 
the vector field $\ell$, and since $\Sigma^i$ is also Lie dragged (in this gauge), all this implies that
\be
\sL_{\ell} (\det(c))=\ell(\det(c))=0.
\ee
Now we must use the rest of the symmetry group of Type I isolated horizons. If we denote by $j_i\in T(H)$ ($i=1,2,3$) the three Killing vectors corresponding to the $SO(3)$ symmetry group of 
Type I isolated horizons. Spherical symmetry of the horizon geometry $(q,D)$ implies \be\sL_{j_i}q=0\ \ \ {\rm and} \ \ \ [\sL_{j_i},D_b] v^a=0\ \ \ \forall v^a\in T(\Delta),\ee
which,through similar manipulations as the one used above, lead to 
 \be j_i (\det{c})=0\ee which completes the prove that $\det{c}$ is constant on $\Delta$.  We can now introduce the dimensionless constant $2\pi c_0:= a_{\va H}\det(c)$. Finally one can fix $c_0=1$ by choosing the appropriate null generator from the equivalence class $[\ell]$. 
$\square$

}

\section{The conserved presymplectic structure}\label{CSS}

In this section we show in detail how the IH boundary condition 
implies  the appearance of an $SU(2)$ Chern-Simons boundary term in the
symplectic structure describing the dynamics of Type I isolated horizons.
This result is key for the quantization of the system described in Section \ref{quentum}.

\subsection{The action principle}
\noindent The conserved pre-symplectic structure in terms of Ashtekar
variables can be easily obtained in the covariant phase space
formalism. The action principle of general relativity in self dual variables containing an inner boundary satisfying the
IH boundary condition (for asymptotically flat spacetimes)
takes the form \be S[e,A_{\va +}]=-\frac{i}{\kappa}\int_{\sM}
\Sigma^{\va +}_i(e)\wedge F^i(A_{\va +}) +\frac{i}{\kappa}\int_{\tau_{\infty}}
\Sigma^{\va +}_i(e)\wedge A^i_{\va +} \label{aacc}\ee where
$\Sigma^{\va +}_i(e)=\epsilon^{i}_{\ jk} e^j\wedge e^k+ i 2 e^0\wedge e^i$ and
$A^i_{\va +}$ is the self-dual connection, and a boundary contribution at a suitable time cylinder $\tau_{\infty}$ at infinity is required 
for the differentiability of the action. {No boundary term is necessary if one allows variations that fix an isolated horizon geometry up to diffeomorphisms and Lorentz  transformations. 
This is a very general property and we shall prove it in the next section as we need a little bit of notation that is introduced there.} 
%Due to the IH boundary condition, no boundary term at the inner boundary is necessary \cite{afk}.

First variation of the
action yields \be \label{fi}\delta S[e,A_{\va +}]=\frac{-i}{\kappa}\int_{\sM}
\delta \Sigma^{\va +}_i(e)\wedge F^i(A_{\va +}) -d_{A_{\va +}} \Sigma^{\va +}_{i}
\wedge \delta A^i_{\va +}+ d( \Sigma^{\va +}_{i} \wedge \delta A^i_{\va +} )+\frac{i}{\kappa}\int_{\tau_{\infty}}
\delta(\Sigma^{\va +}_i(e)\wedge A^i_{\va +}) ,
\ee
from which the self dual version of Einstein's equations follow
\ba
\nonumber && \epsilon_{ijk} e^j\wedge F^i(A_{\va +})+ie^0\wedge F_{k}(A_{\va +})=0\\
\nonumber && e_i\wedge F^i(A_{\va +})=0\\
&& d_{A_{\va +}}\Sigma^{\va +}_i=0,\label{fe}
\ea as the boundary terms in the variation of the action cancel.

\subsection{The classical results in a nutshell}\label{covphase}

In the following subsections a series of technical results are explicitly proven. 
Here we give an account of these results. The reader who is not interested in the explicit proofs can jump directly to 
Section \ref{GSS} after reading the present subsection. 
In this work we study general relativity on a spacetime manifold with an internal
 boundary satisfying the isolated boundary condition corresponding to Type I isolated horizons, and asymptotic 
 flatness at infinity. 
The phase space of such system is denoted
$\Gamma$ and is defined by  an infinite dimensional manifold where points $p\in
\Gamma$ are given by solutions to Einstein's equations satisfying the Type I IH
boundary condition. Explicitly a point $p\in \Gamma$ can be
parametrized by a pair $p = ({\Sigma^{\va +}}, {A_{\va +}})$ satisfying the field
equations (\ref{fe}) and the requirements of Definition \ref{defini}. 
In particular fields at the boundary satisfy Einstein's equations and the 
constraints given in Section \ref{mmain}.
Let ${\rm T_p} (\Gamma)$ denote
the space of variations $\delta=(\delta\Sigma^{\va +}, \delta A_{\va +})$ at $p$ (in symbols $\delta\in {\rm T_p} (\Gamma)$).
A very important point is that the IH boundary conditions
 severely restrict the form of field variations at the horizon.  Thus we have that variations
$\delta=(\delta\Sigma^{\va +},\delta A_{\va +})\in {\rm T_p} (\Gamma)$ are such that
for the pull back of fields on the horizon they correspond to linear
combinations of $SL(2,\C)$ internal gauge transformations and
diffeomorphisms preserving the preferred foliation of $\Delta$. In
equations, for $\alpha: \Delta \rightarrow sl(2,C)$ and $v:
\Delta\rightarrow {\rm T}(H)$ we have that \ba \nonumber \delta
\spb{\Sigma^{\va +}}&=&\delta_{\alpha} \spb{\Sigma^{\va +}}+\delta_v \spb{\Sigma^{\va +}} \\
\delta \spb{A_{\va +}}&=&\delta_{\alpha} \spb{A_{\va +}}+\delta_v
\spb{A_{\va +}} \label{varyvary}\ea where the arrows denote pull-back to $\Delta$, 
and the infinitesimal $SL(2,C)$ transformations
are explicitly given by \be \delta_{\alpha}\Sigma^{\va +}=[\alpha,\Sigma^{\va +}],\ \
\ \delta_{\alpha} A_{\va +}=-d_{A_{\va +}} \alpha, \label{sl2c}\ee while the
diffeomorphisms tangent to H take the following form
\be{\delta_v\Sigma^{\va +}_i=\sL_v\Sigma^{\va +}_i}=
\underbrace{v\inter d_{A_{\va +}}\Sigma^{\va +}_i}_{{\mbox{ \tiny \bf $=0$
(Gauss)}}}+d_{A_{\va +}}(v\inter\Sigma^{\va +})_i{-[v\inter A_{\va
+},\Sigma^{\va +}]_i}\ \ \ \ {\delta_v A_{\va +}^i=\sL_v A_{\va +}^i}=v\inter
F_{\va +}^i{+d_{A_{\va +}}(v\inter A_{\va +})^i},
\label{diffy}\ee 
where $(v\inter \omega)_{b_1\cdots b_{p-1}}\equiv
v^a\omega_{ab_1\cdots b_{p-1}}$ for any $p$-form $\omega_{b_1\cdots
b_p}$, and the first term in the expresion of the Lie derivative of
$\Sigma^{\va +}_i$ can be dropped due to the Gauss constraint $d_A\Sigma^{\va +}_i=0$.

So far we have defined the covariant phase space as an infinite
dimensional manifold. For it to become a phase space it is necessary
to provide it with a presymplectic structure. As the field equations,
the presymplectic structure can be obtained from the first variation
of the action (\ref{fi}). In particular a symplectic potential density
for gravity can be directly read off from the total differential term
in (\ref{fi}) \cite{cov}. The symplectic potential density is
therefore \be \theta(\delta)=\frac{-i}{\kappa}\Sigma^{\va +}_{i} \wedge \delta
A^i_{\va +}\ \ \ \ \forall \ \ \delta\in T_p\Gamma.  \ee and the
symplectic current takes the form \be
J(\delta_1,\delta_2)=-\frac{2i}{\kappa} \delta_{[1}\Sigma^{\va +}_i\wedge \delta_{2]}A_{\va +}^i\ \ \ \ \forall \ \ \delta_1, \delta_2
\in T_p\Gamma. \ee Einstein's equations imply $dJ=0$. Therefore,
applying Stokes theorem to the four dimensional (shaded) region in
Fig. \ref{figui} bounded by $M_1$ in the past, $M_2$ in the future,
a timelike cylinder at spacial infinity  on the right, and the isolated horizon $\Delta$
on the left we obtain \ba \int_{M_1}\!\!\delta_{[1}\Sigma^{\va +}_i\wedge \delta_{2]}A_{\va +}^i - \int_{M_2} \!\!
\delta_{[1}\Sigma^{\va +}_i\wedge \delta_{2]}A_{\va +}^i +
{\int_{\Delta}\!\!  \delta_{[1}\Sigma^{\va +}_i\wedge
\delta_{2]}A_{\va +}^i}=0.\ea Now it turns out that the horizon
integral in this expression is a pure boundary contribution:
the symplectic flux across the horizon can be expressed as a sum of
two terms corresponding to the two-spheres $H_1=\Delta \cap
M_1$ and $H_2=\Delta \cap M_2$. Explicitly (see Proposition 1 proven 
below), the symplectic flux across the horizon $\Delta$ factorizes
into two contributions on $\partial \Delta$ given by $SU(2)$
Chern-Simons presymplectic terms according to \ba
{\int_{\Delta}2\delta_{[1}\Sigma^{\va +}_i\wedge \delta_{2]}A_{\va
+}^i=\frac{a_{\va H}}{2\pi}\left[\int_{H_2}-\int_{H_1}\right]
\delta_{[1}A_{{\va +}i}\wedge \delta_{2]}A^i_{\va +}}. \ea Thus \ba
\int_{M_1}2\delta_{[1}\Sigma^{\va +}_i\wedge \delta_{2]}A_{\va
+}^i-\frac{a_{\va H}}{2\pi}\int_{H_1} \delta_{[1}A_{{\va +}i}\wedge
\delta_{2]}A^i_{\va +} = \int_{M_2}2\delta_{[1}\Sigma^{\va +}_i\wedge
\delta_{2]}A_{\va +}^i-\frac{a_{\va H}}{2\pi}\int_{H_2}
\delta_{[1}A_{{\va +}i}\wedge \delta_{2]}A^i_{\va +} \ea which implies
that the following presymplectic structure is conserved {\be{{i
\kappa\,
\Omega_{M}(\delta_1,\delta_2)=\int_{M}2\delta_{[1}\Sigma^{\va +}_i\wedge \delta_{2]}A_{\va +}^i-\frac{a_{\va H}}{2\pi}\int_{H}
\delta_{[1}A_{{\va +}i}\wedge \delta_{2]}A^i_{\va +}}}
\label{csym},\ee or in other words is independent of $M$. The presence of the boundary term in 
the presymplectic structure might seem at first sight peculiar; however, we
will prove in the following section that the previous symplectic
structure can be written as \be\label{sk}
{\kappa\,\Omega_{M}(\delta_1,\delta_2)=\int_{M}2\delta_{[1}\Sigma_i\wedge
\delta_{2]}K^i}, \ee where we are using the fact that, in the time gauge where $e^0$ is normal to the space slicing, $\Sigma^{\va +i}={\rm Re}[\Sigma^{\va +i}]=\Sigma^i$ when pulled back on $M$. The previous equation is nothing else but the familiar
presymplectic structure of general relativity in terms of the Palatini $\Sigma-K$ variables.  
In essence the boundary term arises when connection variables are used in the parametrization of the gravitational degrees of freedom.

Finally, as shown in Section \ref{barbero}, the key result for the
quantization of Type I IH phase space: the presymplectic structure in
Ashtekar Barbero variables takes the form \ba\label{baba} &&
\!\!\!\!\!\!\!\!\!\!\!\!\!\kappa\beta
\Omega_{M}(\delta_1,\delta_2)=\int_{M}\!\!\! 2\delta_{[1} \Sigma^{i}
\wedge \delta_{2]} A^{\va}_{i}- {\frac{a_{\va H}}{\pi ({1-\beta^2})}}
\int_H \!\!\! \delta_{1} A_i \wedge \delta_2 A^i. \ea
The above equation is the main result of the classical analysis of this paper. It shows that
the conserved presymplectic structure of Type I isolated horizons aquires 
a boundary term given by an $SU(2)$ Chern-Simons presymplectic structure when 
the unconstrained phase space is parametrized in terms of Ashtekar-Barbero variables. 
In the following subsection we prove this equation.

\subsubsection{On the absence of boundary term on the internal boundary}

{Before getting involved with the construction of the conserved presymplectic structure let us come back to the issue
of the differentiability of the action principle. 
In the isolated horizon literature it is argued that the IH boundary condition guaranties the differentiability of the
action principle without the need of the addition of any boundary term (see \cite{afk}). As we show here, this property is satisfied by 
more general kind of boundary conditions.
As mentioned above, the allowed variations are such that  the IH geometry is fixed up to diffeomorphisms of $\Delta$ and gauge transformations.
This enough for the boundary term arising in the first variation of the action (\ref{aacc}) to vanish. 
The boundary term arising on $\Delta$ upon first variation of the action is\be
B(\delta)=-\frac{i}{\kappa}\int_{\Delta} \Sigma_i\wedge\delta A^i
\ee
First let us show that $B(\delta_{\alpha})=0$ for $\delta_{\alpha}$ as given in (\ref{sl2c}). We get
\be
B(\delta_{\alpha})=\frac{i}{\kappa}\int_{\Delta} \Sigma_i\wedge d_A\alpha^i=-\frac{i}{\kappa}\int_{\Delta} (d_A\Sigma_i)\alpha^i-\int_{\partial \Delta} \Sigma_i\alpha^i=0,
\ee
where we integrated by parts in the first identity, the first term in the second identity vanishes due to Eisntein's equations while the
second term vanishes due to the fact that fields are held fixed at the initial and final surfaces $M_1$ and $M_2$ and so $\alpha=0$ when evaluated at $\partial \Delta$.
Similarly we can prove that $B(\delta_{v})=0$ for $\delta_{v}$ as given in (\ref{diffy}) with (this is the only difference) $v\in T(\Delta)$. We get
\ba
&& \nonumber B(\delta_{v})=-\frac{i}{\kappa}\int_{\Delta} \Sigma_i\wedge (v\inter F^i(A)+d_A(v\inter A^i))=\\ &&=-\frac{i}{\kappa}\int_{\Delta} \Sigma_i\wedge (v\inter F^i(A))+\frac{i}{\kappa}\int_{\Delta} d_A\Sigma_i(v\inter A^i)+\int_{\partial \Delta} \Sigma_i(v\inter A^i)=0,
\ea
where in the last line the first and second terms vanish due to Einstein's equations, and the last term vanishes because variations are such that the vector field $v$
vanishes at $\partial \Delta$. Notice that we have not made use of the IH boundary condition.
}

\subsection{The presymplectic structure in self-dual variables} \label{thms}

In this section we prove a series of propositions implying that the
presymplectic structure of Type I isolated horizons is given by equation
(\ref{csym}). In addition, we will prove that the symplectic structure is real and
takes the simple form (\ref{sk}) in terms of Palatini variables.
\vskip.2cm
\noindent {\bf Proposition 1:} The symplectic flux across a Type I isolated horizon $\Delta$ factorizes into boundary contributions at $H_1=\Delta\cap M_1$ and $H_2=\Delta\cap M_2$ according to
 \ba
{\int_{\Delta}\delta_{[1}\Sigma^{\va +}_i\wedge \delta_{2]}A_{\va
+}^i=\frac{a_{\va H}}{2\pi}\left[\int_{H_2}-\int_{H_1}\right]
\delta_{[1}A_{{\va +}i}\wedge \delta_{2]}A^i_{\va +}}. \ea
}
\vskip.2cm

\noindent {\em Proof:}
On $\Delta$ all variations are linear combinations of $SL(2,C)$ transformations and tangent diffeos
as stated in (\ref{varyvary}), (\ref{sl2c}), and (\ref{diffy}).
\[\delta=\delta_{\alpha}+\delta_{v}\]
for $\alpha: \Delta \rightarrow sl(2,C)$ and $v: \Delta\rightarrow
{\rm T}(H)$.  Lets start with $SL(2,\C)$ transformations. Using (\ref{sl2c})
we get \ba \label{gaugy} 
i \kappa\, \Omega_{\Delta}(\delta_\alpha,\delta)&=&\int_{\Delta}
[\alpha,\Sigma]^i\wedge \delta A^{i}_{\va +}+ \delta
\Sigma_i\wedge d_A(\alpha)^i= \int_{\Delta}
-{\alpha_i \delta (d_A\Sigma^i)} + d
(\delta \Sigma_i \alpha^i)=\nonumber \\  \nonumber &=& \int_{\partial\Delta} \delta
\Sigma^i \alpha_i = -\frac{a_{\va
H}}{2\pi}\int_{\partial\Delta} \delta F^i_{\va +} \alpha_i=
-\frac{a_{\va H}}{2\pi}\int_{\partial\Delta} d_A (\delta A^i_{\va +})
\alpha_i= -\frac{a_{\va H}}{2\pi}\int_{\partial\Delta} \delta A^i_{\va
+}\wedge  d_A \alpha_i= \\&=&\frac{a_{\va H}}{2\pi}\int_{\partial\Delta} \delta
A^i_{\va +}\wedge \delta_\alpha A_i, \ea where in the first line we
used the equations of motion $d_A\Sigma^i=0$ and in the second line we used the IH
boundary condition (\ref{unos}). We have therefore shown that \be
\nonumber { i\kappa\, \Omega_{\Delta}(\delta_\alpha,\delta)=-\frac{a_{\va
H}}{2\pi}\int_{\partial \Delta} \delta_\alpha A_{{\va +}i}\wedge \delta
A^{i}_{\va +}} \ee Similarly, for diffeomorphisms we first 
notice that (\ref{diffy}) implies that
\[\delta_v=\delta^{\star}_{v}+\delta_{\alpha(A,v)},\] where
$\alpha(A,v)=v\inter A_{\va +}$ and the explicit form of
$\delta^{\star}_{v}$ is defined as
\[\delta^{\star}_v \Sigma_i=
d_A(v\inter\Sigma)^i,\ \ \ \ \delta^{\star}_v A_{\va
 +}^i=v\inter F_{\va +}^i.\] We have that\ba 
 \nonumber i \kappa \Omega_{\Delta}(\delta^{\star}_v,\delta) &=& \int_{\Delta}
 d_A(v\inter\Sigma)^i\wedge \delta A^{i}_{\va +}-\delta
 \Sigma_i\wedge (v\inter F_{\va +}^i)=\\ \nonumber &=&
 \int_{\Delta} d((v\inter\Sigma)^i\wedge \delta A^{i}_{\va
 +}) +(v\inter\Sigma)^i\wedge d_A(\delta A^{i}_{\va +})-\delta
 \Sigma_i\wedge (v\inter F_{\va +}^i)=\\  \nonumber &=&
 \int_{\Delta} d((v\inter\Sigma)^i\wedge \delta A^{i}_{\va
 +}) {+(v\inter\Sigma)^i\wedge \delta F^{i}_{\va
 +}-\delta \Sigma_i\wedge (v\inter F_{\va
 +}^i)}=\int_{\Delta} d((v\inter\Sigma)^i\wedge \delta A^{i}_{\va
 +}) + \delta (\Sigma_i[v\inter F^i(A_{\va +})])\\ &=& \int_{\partial \Delta} (v\inter\Sigma_{\va
 +})^i\wedge \delta A^{i}_{\va +}= -\frac{a_{\va
 H}}{2\pi}\int_{\partial \Delta} \delta^{\star}_v A_{\va +}^i\wedge \delta
 A^{i}_{\va +}, \ea
where in the third line we used the vector constraint $\Sigma_i[v\inter F^i(A_{\va +})]=0$, while in last line we have used the equations of motion and equation (\ref{unos}).
Notice that the calculation leading to equation (\ref{gaugy}) is also valid for a field dependent $\alpha$
such as $\alpha(A,v)$. This plus the linearity of the presymplectic structure lead to
\be
 {{i\kappa\, \Omega_{\Delta}(\delta_v,\delta)=-\frac{a_{\va
H}}{2\pi}\int_{\partial \Delta} \delta_v A_{{\va +}i}\wedge \delta
A^{i}_{\va +}}}
\ee
and concludes the proof of our proposition $\square$.

\vskip.2cm
The previous proposition implies that the presymplectic structure (\ref{csym}) is indeed conserved by evolution in $\Gamma$.
Now we are ready to state the next important proposition.
\vskip.2cm

\noindent{\bf Proposition 2:} The presymplectic form $\Omega_{M}(\delta_1,\delta_2)$ given by
\[{{i\kappa\, \Omega_{M}(\delta_1,\delta_2)=\int_{M}2 \delta_{[1}\Sigma^{\va +}_i\wedge
\delta_{2]}A_{\va +}^i-\frac{a_{\va H}}{2\pi}\int_{H}
\delta_{[1}A_{{\va +}i}\wedge \delta_{2]}A^i_{\va +}}}
\] is independent of $M$ and {\em real}. Moreover, the symplectic structure can be described entirely in terms of variables $K\equiv Im(A_{\va +})$
and $\Sigma$ taking the familliar form
\be {\kappa\,\Omega_M(\delta_1,\delta_2)=\int_{M}2\delta_{[1}\Sigma_i\wedge
\delta_{2]}K^i},\label{noboundary}\ee
which is manifestly real and has no boundary contribution.
\vskip.2cm

\noindent {\it Proof:} The independence of the sysmplectic structure on $M$ follows directly from Proposition 2 and the argument presented at the end of the previous section. Now let us analyze the reality of the presymplectic structure. The symplectic potential for $\Omega$ written in terms of self dual variables
is
\be{{i\, \kappa\ \Theta(\delta)=\int_{M} \Sigma_i\wedge
\delta A_{\va +}^i -\frac{a_{\va H}}{4\pi}\int_{H}
A_{{\va +}i}\wedge \delta A^i_{\va +}}}
\ee
Using $A_{+}^i=\Gamma^i+iK^i$ we get 
\be {{\kappa\ \Theta(\delta)=\int_{M} \Sigma_i\wedge
\delta K^i- i\left( \int_{M} \Sigma_i\wedge
\delta \Gamma^i - \frac{a_{\va H}}{4\pi}\int_{H}
A_{{\va +}i}\wedge \delta A^i_{\va +} \right)}}\ee
Using a well known property of the spin connection \cite{thiemann}, and denoting by $\theta_{0}(\delta)$ the term in parenthesis in the previous equation, we have\ba &&\nonumber 
 \Theta_0(\delta)\equiv \int_{M} \Sigma_i\wedge
\delta \Gamma^i - \frac{a_{\va H}}{4\pi}\int_{H}
A_{{\va +}i}\wedge \delta A^i_{\va +} = \int_{H} - e_i \wedge
\delta e^i - \frac{a_{\va H}}{4\pi}\int_{H}
A_{{\va +}i}\wedge \delta A^i_{\va +}
\ea
The proposition follows from that fact that $\Theta_0(\delta)$ vanishes as proven in the following Lemma $\square$. 

\vskip.2cm
{\noindent \bf Lemma 3: } The phase space one-form $\Theta_0(\delta)$ defined by \be
\Theta_0(\delta) \equiv  \int_{H} - e_i \wedge
\delta e^i - \frac{a_{\va H}}{4\pi}\int_{H}
A_{{\va +}i}\wedge \delta A^i_{\va +}
\ee is closed.

\vskip.2cm
{\noindent \em Proof:} From the definition of the phase space $\Gamma$ given in Section \ref{covphase}, in particular from Eqs. (\ref{varyvary}) we know  that \ba \nonumber
\delta \spb{e}&=&\delta_{\alpha} \spb{e}+\delta_v \spb{e} \\ \delta
\spb{A_{\va +}}&=&\delta_{\alpha} \spb{A_{\va +}}+\delta_v \spb{A_{\va
+}}. \ea Let us denote by 
\[{\frak d}\Theta_0 (\delta_1,\delta_2)=\delta_1(\Theta_0(\delta_{2}))-\delta_2 (\Theta_0(\delta_{1}))\]
the exterior derivative of $\Theta_0$.  For infinitesimal $SL(2,C)$ transformations we have \be
\delta_{\alpha}e=[\alpha,e],\ \ \ \delta_{\alpha} A=-d_A \alpha,
\label{sl2cc}, \ee from which it follows \ba \nonumber {\frak d}\Theta_0(\delta, \delta_{\alpha})&=&
\int_{H} - 2 \delta e_i \wedge [\alpha,e]^i + \frac{a_{\va H}}{2\pi} \delta A_{{\va
+}i}\wedge d_{A}\alpha^i=\int_{H} - 2 \delta e_i \wedge [\alpha,e]^i +
\frac{a_{\va H}}{2\pi} \delta F^i(A_{{\va +}}) \alpha_i=\\ &=& \int_{H} \delta (e^j
\wedge e^k) \alpha^i\epsilon_{ijk} + \frac{a_{\va H}}{2\pi} \delta F^i(A_{{\va
+}}) \alpha_i= \int_{H} \delta[\Sigma^i + \frac{a_{\va H}}{2\pi} F^i(A_{{\va
+}})] \alpha_i=0, \ea where in the first line we have integrated by
parts, and in the second line we used the IH boundary condition.  The action of
diffeomorphisms tangent to H on the connection and triad take the following form \be{\delta_v
e^i=\sL_v e^i=d(v\inter e^i)+v\inter de^i}\ \ \ \ {\delta_v A_{\va
+}^i=\sL_v A_{\va +}^i=v\inter F^i(A_{\va +})+d_{A_{\va +}} (v\inter
A_{\va +}^i)}.\label{diffyc}\ee Now we have \ba \nonumber
{\frak d}\Theta_0(\delta, \delta_{v})&=& \int_{H} -2 \delta e_i \wedge \sL_v e^i-\frac{a_{\va
H}}{2\pi } \delta A_{{\va +}i}\wedge \sL_v A_{\va +}^i=\\ \nonumber &=&
\int_{H} -2\delta e_i \wedge d(v\inter e^i) - 2\delta e_i \wedge v\inter de^i-
\frac{a_{\va H}}{2\pi } [\delta A_{{\va +}i}\wedge v\inter F^i(A_{\va +})+
\delta A_{{\va +}i}\wedge d_{A_{\va +}} (v\inter A_{\va +}^i)]=\\ &=&
\int_{H} -2\delta e_i \wedge d(v\inter e^i) - 2 v\inter \delta e_i \wedge de^i - 
\frac{a_{\va H}}{2\pi } [\delta (v\inter A_{{\va +}i})\wedge F^i(A_{\va +})+
\delta F_i(A_{{\va +}})\wedge v\inter A_{\va +}^i]
\nonumber \\ &=& \int_{H}-2\delta (de_i) \wedge v\inter e^i - 2 v\inter \delta e_i \wedge de^i - \frac{a_{\va H}}{2\pi
} \delta [v\inter A_{{\va +}i}\wedge F^i(A_{\va +})]= \nonumber \\ &=&
\int_{H}\delta[v\inter \Gamma^i \wedge \Sigma_i- v\inter (\Gamma^i+i
K^i)\wedge \Sigma_i]=0, \ea where in addition to integrating by parts
and using that $\partial H=0$, we have used the identity $A\wedge
(v\inter B)-(v\inter A)\wedge B=0$ for $A$ a $1$-form and $B$ a
$2$-form in a two dimensional manifold, and Cartan's structure equation $de^i+\epsilon_{ijk} \Gamma^j e^k=0$. In the 
last line we used eq. (\ref{unos}), and eq. (\ref{doss})---which implies
that $K^i\Sigma_i=0$ $\square$.

\subsection{Presymplectic structure in Ashtekar-Barbero variables}\label{barbero}

In the previous section (Proposition 2) we have shown how the presymplectic structure
\begin{equation}
\label{sylstr1}
\Omega_{M}(\delta_1,\delta_2)=\frac{1}{\kappa}\int_{M} [\delta_1 \Sigma^{i} \wedge \delta_2
K^{\va}_{i}-\delta_2 \Sigma^{i} \wedge \delta_1 K^{\va}_{i}]
\end{equation}
is indeed preserved in the presence of an IH. More precisely in the shaded
space-time region in Fig. \ref{figui} one has \be
\Omega_{M_2}(\delta_1,\delta_2)= \Omega_{M_1}(\delta_1,\delta_2).\ee
That is, the symplectic flux across the isolated horizon $\Delta$
{\em vanishes}  due to the isolated horizon boundary condition
\cite{ack}.  We will show now, how the very same presymplectic structure
takes the form (\ref{baba}) when written in terms of the Ashtekar-Barbero connection variables.
For this we need to prove the following lemma:

\vskip.2cm
{\noindent \bf Lemma 4: } The phase space one-form $\Theta^{\beta}_0(\delta)$ defined by \be
\beta \Theta^{\beta}_0(\delta) \equiv  \int_{H} - e_i \wedge
\delta e^i - \frac{a_{\va H}}{2\pi(1-\beta^2)}\int_{H}\label{58}
A_i\wedge \delta A^i
\ee is closed.

\vskip.2cm
{\noindent \em Proof:}
from the definition of the phase space (Section \ref{covphase}) we have
\ba \nonumber \delta
\spb{e}&=&\delta_{\alpha} \spb{e}+\delta_v \spb{e} \\
\delta \spb{A_{\va +}}&=&\delta_{\alpha} \spb{A_{\va +}}+\delta_v
\spb{A_{\va +}}. \ea Let us first study $\Theta^{\beta}_0(\delta_{\alpha})$ where the infinitesimal $SU(2)$ transformation
are explicitly given by \be \delta_{\alpha}e=[\alpha,e],\ \
\ \delta_{\alpha} A=-d_A \alpha, \label{su2}.\ee 
We have \ba \nonumber
 \beta {\frak d}\Theta^{\beta}_0(\delta, \delta_{\alpha})&=& \int_{H} -2 \delta e_i \wedge
[\alpha,e]^i + \frac{a_{\va H}}{\pi(1-\beta^2)}
\delta A_{i}\wedge d_{A}\alpha^i=\int_{H} - 2\delta e_i \wedge
[\alpha,e]^i + \frac{a_{\va H}}{\pi (1-\beta^2)}
\delta F^i(A) \alpha_i=\\ &=&
\int_{H} \delta [\Sigma^i + \frac{a_{\va H}}{\pi(1-\beta^2)}
F^i(A)] \alpha_i=0,
\ea
where in the first line we have integrated by parts, and in the second line we used 
the IH boundary condition. The proof that the presymplectic potential vanishes for $\delta_v$ mimics exactly the corresponding part of the proof of Lemma 3 $\square$.

\vskip.2cm

The next step is to write the symplectic structure in terms of
Ashtekar-Barbero connection variables necessary for the quantization
program of LQG. When there is no boundary the $SU(2)$ connection \be
A_a^i=\Gamma^i+ {\beta} K_a^i\ee is canonically conjugate to
$\epsilon^{abc}\beta^{-1} \Sigma_{bc}^i/4$ where $\beta$ is the
so-called Immirzi parameter. In the presence of a boundary the
situation is more subtle: the symplectic structure acquires a
boundary term. 
\vskip.2cm

\noindent{\bf Proposition 3:}  In terms of Ashtekar-Barbero variables the presymplectic structure
of the spherically symmetric isolated horizon takes the form
\ba
\label{BIsylstr} && \!\!\!\!\!\!\!\!\!\!\!\!\!\kappa\beta \Omega_{M}=\int_{M}\!\!\! 2\delta_{[1}
\Sigma^{i} \wedge \delta_{2]} A^{\va}_{i}- {\frac{a_{\va H}}{\pi
({1-\beta^2})}} \int_H \!\!\! \delta_{1} A_i \wedge \delta_2
 A^i, \ea
where $\kappa={32\pi G}$.  

\vskip.2cm
\noindent{\em Proof:}
The result follows from the variation of the presymplectic potential 
\ba && \nonumber \kappa \beta \Theta(\delta) =\int_{M} \Sigma_i\wedge (\beta \delta K^i)+\beta\Theta_0^{\beta}(\delta)=\\ \nonumber && 
= \int_M \Sigma_i\wedge \delta (\Gamma^i +\beta K^i)-{\frac{a_{\va H}}{2\pi
({1-\beta^2})}} \int_H \!\!\! A_i \wedge \delta
 A^i,\label{ssqq}\ea
which is simply the presymplectic potential leading to the conserved presymplectic structure in $\Sigma-K$ variables (in equation (\ref{noboundary})) to which we have added a term proportional to $\Theta^{\beta}_0(\delta)$; a closed term which does not affect the presymplectic structure according to Lemma 4 $\square$.
\vskip.2cm
{ \noindent {\em Remark:} Notice that one could have introduced a new connection $\bar A^i=\Gamma^i+\bar \beta K^i$ with a new parameter $\bar\beta$ independent of the Immirzi parameter. The  statement of the
previous lemma would have remained true if on the right hand side of equation (\ref{58}) one would have replaced $\beta$ by $\bar \beta$ and $A^i$ by $\bar A^i$. Consequently, the presymplectic structure can also be parametrized in terms of
the analog of equation (\ref{BIsylstr}) with a boundary term where $A^i$ is replaced by $\bar A^i$ and $\beta$ on the prefactor of the boundary term is replaced by $\bar\beta$. This implies that the description of the boundary 
term in terms of Chern-Simons theory allow for the introduction of a new independent parameter $\bar\beta$ which has the effect of modifying the Chern-Simons level. This ambiguity in the description of the boundary 
degrees of freedom has however no effect in the value of the entropy.
}
%
% Wasn't sure if I should add this:
%
% and using in part gauge-fixed identities
% in \cite{ack}.}

\subsection{A side remark on the triad as the boundary degrees of freedom}\label{sidy}
{
Here we show that one can write the presymplectic structure
\begin{equation}
\Omega_{M}(\delta_1,\delta_2)=\frac{1}{\kappa}\int_{M} [\delta_1 \Sigma^{i} \wedge \delta_2
K^{\va}_{i}-\delta_2 \Sigma^{i} \wedge \delta_1 K^{\va}_{i}]
\end{equation}
in a way such that a surface term depending only on the pull back of the triad field while 
the bulk term coincides with the one obtained in the previous section in terms of real connection variables.
In order to do this we rewrite the symplectic potential as follows:
\ba\nonumber 
 \kappa \beta\Phi(\delta)&=&\int_M \Sigma_i\wedge \delta(\beta K^i)\\ \nonumber &=&\int_M \Sigma_i\wedge \delta(\beta K^i+\Gamma^i)-\int_{H}
\Sigma_i\wedge \delta\Gamma^i\\ &=&\int_M \Sigma_i\wedge \delta A^i+\int_{H}
e_i\wedge \delta e^i.
\ea 
As a result the symplectic structure becomes \cite{noni} (and independently \cite{wi})
\begin{equation} \label{ppp}
\Omega_{M}(\delta_1,\delta_2)=\frac{1}{\kappa\beta}\int_{M} [\delta_1 \Sigma^{i} \wedge \delta_2
A^{\va}_{i}-\delta_2 \Sigma^{i} \wedge \delta_1 A^{\va}_{i}]+\frac{2}{\beta \kappa} \int_{H} \delta_{1}e_i\wedge\delta_{2}e^i.
\end{equation}
The previous equation shows that the boundary degrees of freedom could be described in terms of 
the pull back of the triad on the horizon. One could try to quantize the IH system in this formulation in order to address the
question of black hole entropy calculation. Such project would be certainly interesting. However, the treatment is clearly not immediate as it would require 
the background independent quantization of the triad field on the boundary for which the usual available techniques do not seem to naturally apply.  
Nevertheless,  the previous equations provides an interesting insight already at the classical level, as the boundary symplectic structure, written in this way, has a remarkable implication
for geometric quantities of interest in the first order formulation. To see this let us take $S\subset H$ and $\alpha: H\to su(2)$ so that we can introduce the fluxes $\Sigma (S,\alpha)$ according to
\be
\Sigma (S,\alpha)=\int_{S\subset H} {\rm Tr}[\alpha \Sigma],
\ee
where $ {\rm Tr}[\alpha \Sigma]=\epsilon_{ijk}\alpha^i e^j\wedge e^k$. Now (\ref{ppp}) implies the Poisson bracket $\{e^i_a(x),e^j_b(y)\}=\epsilon_{ab}\delta^{ij} \delta(x,y)$ from which the following remarkable equation follows:
\be\label{hfa}
\{\Sigma (S,\alpha),\Sigma (S',\beta) \}=\Sigma (S\cap S',[\alpha,\beta]). 
\ee 
The Poisson brackets among surface fluxes is non vanishing and reproduces the $su(2)$ Lie algebra!
This is an interesting property that we find entirely in terms of classical considerations using 
smooth field configurations. 
However, compatibility with the bulk fields seems to single out the treatments of kinematical degrees of freedom 
in terms of the so called {\em holonomy-flux} algebra of classical observables for which flux variables 
satisfy the exact analog of (\ref{hfa}) as described in \cite{zapata}. This fact strengthens  
even further the relevance of the uniqueness theorems \cite{lost}, as they assume the use of the {\em holonomy-flux} algebra as the starting point for quantization.
}

\section{Gauge symmetries}\label{GSS}

In this section we rederive the form of the presymplectic symplectic structure written in Ashtekar-Barbero variables
by means of gauge symmetry argument. The idea is to first study the gauge symmetries of the presymplectic structure when written 
in Palatini variables, as in Equation (\ref{sk}). We will show that, due to the nature of variations at the horizon, the boundary term
in Equation (\ref{baba}) is completely fixed by the requirement that the gauge symmetry content is unchanged when the presymplectic structure is parametrized by  
Ashtekar-Barbero variables. This argument is completely equivalent to the content of the previous section
and was used in \cite{nous} as a shortcut construction of the presymplectic structure for Type I isolated horizons
in terms of real connection variables. Another important result of this section is the computation of the classical constraint algebra
in Subsection \ref{CCA} which are essential for clarifying the quantization strategy implemented in Section \ref{quentum}

The gauge symmetry content of the phase space $\Gamma$ is implied by the following proposition.

\vskip.2cm
\noindent{\bf Proposition 4:}  Phase space
tangent vectors $\delta_\alpha, \delta_{v}\in T_p\Gamma$ of the form
\begin{eqnarray} &&\nonumber 
\delta_{\alpha} \Sigma=[\alpha,\Sigma],\ \
\delta_{\alpha}K=[\alpha,K]; \\ && 
\delta_{v} \Sigma= \Lie_v \Sigma=v\inter d\Sigma+d(v\inter\Sigma), \
\ \delta_{v}K= \Lie_v K=v\inter dK+d(v\inter K)
\label{ttta}\end{eqnarray}
for $\alpha: M \rightarrow \frak{su}(2)$ and $v \in \Vect(M)$
tangent to the horizon, are degenerate directions of $\Omega_{M}$.

\vskip.2cm
\noindent {\em Proof:} The proof follows from manipulations very similar in spirit to the ones used for proving the previous propositions.
We start with the $SU(2)$ transformations $\delta_\alpha$, and 
we get \ba \label{su2} &&
\kappa \Omega_{M}(\delta_\alpha,\delta)=\int_{M}
[\alpha,\Sigma]_i\wedge \delta K^{i}- \delta
\Sigma_i\wedge [\alpha,K]^i= \int_{M}
\delta({\epsilon_{ijk}\alpha^j \Sigma^k\wedge K^i)})=0, \ea where we
used the Gauss constraint $\epsilon_{ijk}\Sigma^k\wedge K^i=0$. In order to treat the case of the infinitesimal diffeomorphims tangent to the horizon $H$ it will be convenient to first write the form of the vector constraint $V_a$ in terms of $\Sigma-K$ variables \cite{tate}. We have
\ba
v\inter V=dK^i\wedge v\inter\Sigma_i+v\inter K^i\ d\Sigma_i\approx0
\ea
variations of the previous equation yields
\ba
v\inter \delta V&=&d(\delta K)^i \wedge v\inter\Sigma_i+dK^i\wedge v\inter\delta\Sigma_i+v\inter \delta K^i\ d\Sigma_i+v\inter K^i\ d(\delta\Sigma)_i=\nonumber \\ &=&  v\inter\Sigma_i\wedge d(\delta K)^i-\delta\Sigma_i\wedge v\inter dK^i+  v\inter d\Sigma_i \wedge \delta K^i+ d(\delta\Sigma)_iv\inter K^i =0,
\ea
where in the second line we have put all the $K$'s to the right, and modified the second and third terms using the identities $A\wedge (v\inter B)+(v\inter A)\wedge B=0$ that is valid for any two $2$-forms $A$ and $B$ on a $3$-manifold, and $A\wedge (v\inter B)-(v\inter A)\wedge B=0$ for a $1$-form $A$ and a $3$-form $B$ on a $3$-manifold respectively. We are now ready to show that $\delta_v$ is a null direction of $\Omega_M$. Explicitly:
\ba &&\nonumber 
\kappa \Omega_{M}(\delta_v,\delta)=\int_{M}
(v\inter d\Sigma+d(v\inter\Sigma))_i\wedge \delta K^{i}- \delta
\Sigma_i\wedge (v\inter dK+d(v\inter K))^i=\\
&&\nonumber =  \int_{M}
v\inter d\Sigma_i\wedge \delta K^{i}+d(v\inter\Sigma)_i\wedge \delta K^{i}- \delta
\Sigma^i\wedge v\inter dK_i- \delta
\Sigma_i \wedge d(v\inter K)^i=\\
&&\nonumber =  \int_{M}\underbrace{
v\inter d\Sigma_i\wedge \delta K^{i}+v\inter\Sigma_i\wedge d(\delta K)^{i}- \delta
\Sigma^i\wedge v\inter dK_i+d(\delta
\Sigma)^i \wedge v\inter K_i}_{v\inter \delta V=0}+\\ && 
+\int_{\partial M}v\inter\Sigma_i\wedge \delta K^{i} -\delta
\Sigma_i \wedge v\inter K^i=\int_{\partial M}\delta(v\inter\Sigma_i\wedge K^{i})=0,  \label{diffh}\ea
where in the last line we have used the identity $v\inter A\wedge B+A\wedge v\inter B=0$ valid for an 
arbitrary $2$-form $A$ and arbitrary $1$-form $B$ on a $2$-manifold, the fact that $v$ is tangent to $H$, and the IH boundary condition Eq. (\ref{doss})
implying $\Sigma_i \wedge K^i=0$ when pulled back on $H$ $\square$.

\vskip.2cm

The previous proposition shows that the IH boundary condition breaks neither the
symmetry under diffeomorphisms preserving $H$ nor the $SU(2)$ internal gauge
symmetry introduced by the use of triad variables.

The gauge invariances of the IH system have been made explicit in the
$\Sigma-K$ parametrization of the presymplectic structure.  However,
due to the results of Propositions 2 and 3, these can also be made
explicit in the parametrization of the presymplectic structure using
either self-dual connection variables or real Ashtekar-Barbero
variables. It is in fact possible to uniquely determine the horizon
contributions to the presymplectic structure in connection variables
entirely in terms of the requirement the transformations (\ref{ttta})
be gauge invariances of the standard bulk presymplectic contribution
plus a suitable boundary term.  More precisely, the requirement of
$SU(2)$ local invariance becomes \be 0=\kappa\beta
\Omega_M(\delta_\alpha,\delta)=\int_{\va M}\!\!\!  \delta_{\alpha}
\Sigma_i \wedge \delta A^i\! - \! \delta \Sigma_i \wedge
\delta_{\alpha} A^i\!+\kappa \beta \Omega_H\ \ \ \ \forall \ \ \ \delta\in
{\rm T_p}(\Gamma), \label{puget} \ee for an (in principle) unknown horizon
contribution to the presymplectic structure $\Omega_H$.  This gives
some information about the nature of the boundary term, namely
\begin{eqnarray*}&&
-\kappa\beta \Omega_{\va H}\!=\! \int_{\va M}\!\!\!  \delta_{\alpha}
\Sigma_i \wedge \delta A^i\! - \! \delta \Sigma_i \wedge
\delta_{\alpha} A^i\!=\!
 \int_{\va M}\!\!\! [\alpha,\Sigma]_i \wedge \delta A^i \!+ \!\delta \Sigma_i \wedge d_A \alpha^i
\\ &&
\!=\!\int_{\va M}\!\!\! d (\alpha_i \delta\Sigma^i)\!-\!\alpha_i
\delta(d_A \Sigma^i)\! =\! - \frac{a_{\va H}}{\pi (1-\beta^2)}
\int_{\va H} \alpha_i \delta F^i(A) \\ && \!=\!
 \frac{a_{\va
H}}{\pi (1-\beta^2)} \int_{\va H} \delta_{\alpha} A_i
\wedge \delta
 A^i.
\end{eqnarray*} where we used the Gauss law
$\delta (d_A\Sigma)=0$, condition (\ref{tress}), and that boundary
terms at infinity vanish. A similar calculation for diffeomorphisms
tangent to the horizon gives an equivalent result. This together with
the nature of variations at the horizon (see Eqs. (\ref{varyvary})) provides an independent way of
establishing the results of Proposition 3. This alternative derivation
of the conserved presymplectic structure was used in \cite{nous}.

\subsection{On the first class nature of the IH constraints}\label{CCA}
{The previous equation above can be written as
\be
\kappa \beta \Omega(\delta_{\alpha}, \delta)=-\int_M \alpha_i \delta(d_A \Sigma^i)-\int_{H} \alpha_i \left[\frac{a_{\va H}}{\pi (1-\beta^2)}\delta F^i+\delta \Sigma^i\right],
\ee
or equivalently
\be
\Omega(\delta_{\alpha}, \delta)+\delta G[\alpha, A, \Sigma]=0,\label{hamil}
\ee
where
\be
G[\alpha, A, \Sigma]=\int_M  \alpha_i (d_A \Sigma^i/(\kappa\beta))+ \int_{H} \alpha_i \left[\frac{a_{\va H}}{\pi \kappa\beta (1-\beta^2)} F^i+ \frac{1}{\kappa\beta}{\Sigma^i}\right].
\ee
In the canonical framework Equation \ref{hamil} implies that local $SU(2)$ transformations $\delta_{\alpha}$ are Hamiltonian vector fields
generated by the ``Hamiltonian" $G[\alpha, A, \Sigma]$. It follows immediately from the definition of Poisson brackets that the Poisson algebra
of $G[\alpha, A, \Sigma]$ closes. More precisely,  one has
\be
\{G[\alpha,  A, \Sigma],G[\beta, A, \Sigma]\}=\Omega(\delta_{\alpha},\delta_{\beta})=\delta_{\beta} G(\alpha, A, \Sigma)
\ee
from where we get 
\be
\{G[\alpha,  A, \Sigma],G[\beta, A, \Sigma]\}=
G([\alpha,\beta], A, \Sigma)\label{algebra}.
\ee
Not surprisingly we get the $SU(2)$ Lie algebra a local $SU(2)$ transformations. In the previous section we showed that these local transformations are indeed gauge transformations.
This implies, in the canonical picture, that canonical variables must satisfy the constraints
\be
G(\alpha,A,\Sigma)\approx 0\ \ \ \forall \ \ \ \alpha:H\cup M \to su(2).
\ee
Now let us look at diffeomorphisms.
A calculation based on the analog of equation (\ref{puget}) for an infinitesimal diffeormorphism preserving $H$ yields
\be
\Omega(\delta_{v}, \delta)+\delta V[v, A, \Sigma]=0,\label{doffi}
\ee 
where
\be
V[v, A, \Sigma]=\int_M  \frac{1}{\kappa\beta}\left[\Sigma_i\wedge v\inter F^i -v\inter A_i d_A\Sigma^i \right]- \int_{H} v\inter A_i \left[\frac{a_{\va H}}{\pi \kappa\beta (1-\beta^2)} F^i+ \frac{1}{\kappa\beta}{\Sigma^i}\right].
\ee
 Finally, a simple calculation as the one leading to (\ref{algebra}), leads to the following first-class constraint algebra
 \ba\n && \{G[\alpha,  A, \Sigma],G[\beta, A, \Sigma]\}=
G([\alpha,\beta], A, \Sigma)\\ \n && 
 \{G[\alpha,  A, \Sigma],V[v, A, \Sigma]\}=G(\sL_v \alpha, A, \Sigma) \\ &&
  \{V[v,  A, \Sigma],V[w, A, \Sigma]\}=V([v,w], A, \Sigma),
 \ea
 where we have ignored the Poisson brackets involving the scalar constraint\footnote{Recall that the smearing of the scalar constraints must vanish on $H$ and hence the full constraint algebra including the scalar constraint will remain first class.}.  Using $\alpha$ and $v$ with support only on the horizon $H$ we can now conclude that the IH boundary condition is first class which justifies the
Dirac implementation that will be carried our in the quantum theory.}

\section{The zeroth and first laws of BH mechanics for (spherical) isolated horizons }\label{firstlaw}

The definition given in Section \ref{defini} implies authomaticaly the
zeroth law of BH mechanics as $\kappa_{\va H}$ is constant on $\Delta$.
In turn, the first law cannot be tested unless a definition of energy
of the IH is given. Due to the fully dynamical nature of the
gravitation field in the neighbourhood of the horizon this might seem
problematic.  Of course one can in addition postulate an energy
formula for the IH in order to satisfy {\em de facto} the first
law. Fortunately, there is a more elegant way. This consists in
requiring the time evolution along vector fields $t^a\in {\rm T}(\sM)$ which
are time translations at infinity and proportional to the null
generators $\ell$ at the horizon to correspond to a Hamiltonian time
evolution \cite{afk}. More precisely, denote by $\delta_t: \Gamma \rightarrow T(\Gamma)$
the phase space tangent vector field associated to an infinitesimal time evolution
along the vector field $t^a$ (which we allow to depend on the phase space point). 
Then $\delta_t$ is Hamiltonian if there exists a functional $H$
such that \be\delta H=\Omega_M(\delta,\delta_t)
\label{Hcondition}
\ee
This requirement  fixes a family of good energy formula and translates into the first law of isolated horizons\be \delta E_{\va
H}=\frac{\kappa_{\va H}}{\kappa} \delta a_{\va H}+\Phi_{\va H} \delta
Q_{\va H}+\mbox{other work terms}, \ee where we have put the explicit
expression of the electromagnetic work term where $\Phi_{\va H}$ is
the electromagnetic potential (constant due to the IH boundary
condition) and $Q_{\va H}$ is the electric charge. 
The above equation implies that $\kappa_{\va H}$ and $\Phi_{\va H}$
to be functions of the IH area $a_{H}$ and charge $Q_{\va H}$ alone.
A unique energy formula is singled out if
we require $\kappa_{\va H}$ to coincide with the
surface gravity of Type I stationary BHs, i.e., those in the Reissner-Nordstrom 
family:
\be
\kappa_{\va H}=\frac{\sqrt{(M^2-Q^2)}} {2M[M+\sqrt{(M^2-Q^2)}]-Q^2}.
\ee

Here we can
explicitly prove the above statement in terms of our variables. We shall make here the simplifying assumption that
there are no matter fields, i.e. , we work in the vacuum case. The explicit form of $\delta_t$ is given by 
\ba\nonumber \delta_t\Sigma&=&\sL_t\Sigma=t \inter
d\Sigma+d(t \inter\Sigma)\\ \delta_t K&=&\sL_tK=t \inter dK+d(t \inter
K).\ea We can now explicitly write the main condition, namely \ba
\label{difft} &&\nonumber 16 \pi G\
\Omega_{M}(\delta_t,\delta)=\int_{M} (t \inter d\Sigma+d(t
\inter\Sigma))_i\wedge \delta K^{i}- \delta \Sigma_i\wedge (t\inter
dK+d(t\inter K))^i=\\ &&\nonumber = \int_{M} t\inter d\Sigma_i\wedge
\delta K^{i}+d(t\inter\Sigma)_i\wedge \delta K^{i}- \delta
\Sigma^i\wedge t\inter dK_i- \delta \Sigma_i \wedge d(t\inter K)^i=\\
\nonumber &&= \int_{\partial M}\ell \inter\Sigma_i\wedge \delta K^{i}
-\delta \Sigma_i \wedge \ell\inter K^i=- \int_{\partial M} \delta
\Sigma_i \wedge \ell\inter K^i=\\ && = 2 \kappa_{\va H}\ \delta a_{\va
H}+\delta E_{\va ADM},\ea Where we have used the same kind of
manipulations used in equation (\ref{diffh}) paying special attention to the fact that the relevant vector field $t=\ell$ 
is (at the horizon) no longer tangent to the horizon cross-section, and
 the fact that the first term in the
third line vanishes due to the IH boundary condition \footnote{This
follows from equations (\ref{doss}), (\ref{sigg}), and (\ref{capa})
implying that \ba&& \nonumber \ell \inter\Sigma_i\wedge \delta
K^{i}=-e^{\alpha} e^3\wedge \delta \left({
e^2}{\sqrt{\frac{2\pi}{a_{\va H}}}}\right)+e^{\alpha} e^2\wedge \delta
\left({ e^3}{\sqrt{\frac{2\pi}{a_{\va H}}}}\right) =\\
\nonumber && =e^{\alpha} \left(\sqrt{\frac{2\pi}{a_{\va
H}}}\delta \left(e^2\wedge e^3\right) +2 e^2\wedge
e^3\delta\left(\sqrt{\frac{2\pi}{a_{\va H}}}\right)\right).\ea Integrating the
previous expression on the horizon gives zero.}.

The condition $\delta H_t =\Omega_M(\delta_t, \delta)$ is solved by $H_t=E_{\va ADM}-E_{\va H}$
with \be
\delta E_{\va H}=\frac{\kappa_{\va H}}{\kappa} \delta a_{\va H}.
\ee
Demanding time evolution to be Hamiltonian singles out a notion of isolated horizon energy which automatically satisfies, 
by this requirement, the first law of black hole mechanics (now extended from the static or locally static context to the isolated horizon context). The general treatment and derivation of the first law can be found in \cite{afk,abl2001}.

\section{Quantization}\label{quentum}

{The form of the symplectic structure motivates one to handle the
quantization of the bulk and horizon degrees of freedom (d.o.f.)
separately.
We first discuss the bulk quantization. As in standard LQG [8] one
first considers (bulk) Hilbert spaces $\sH^B_\gamma$ defined on a
graph $\gamma \subset M$ and then takes the projective limit
containing the Hilbert spaces for arbitrary graphs. Along these
lines let us first consider $\sH^{\va B}_{\gamma}$ for a fixed graph
$\gamma \subset M$ with end points on $H$, denoted $\gamma\cap H$.
The quantum operator associated with $\Sigma$ in (\ref{tress}) is
%
%IMPORTANT: IN THE EQUATION BELOW I AM USING THAT \EPSILON\SIGMA=4E AND THAT THE FLUX IS HALF DUE TO THE FACT THAT H IS A BOUNDARY OF SPACE
%
\begin{equation}
\label{gammasigma} \epsilon^{ab}\hat{\Sigma}^i_{ab}(x) = 16 \pi G
\beta \sum_{p \in \gamma\cap H} \delta(x,x_p) \hat{J}^i(p)
\end{equation}
where $[\hat{J}^i(p),\hat{J}^j(p)]=\epsilon^{ij}_{\ \ k} \hat{J}^k(p)$ at each $p\in\gamma\cap H$.
%
% and $\epsilon^{ab}$ is the density weight 1 Levi-Civita tensor on $S$
%
Consider a basis of $\sH^{{\va B}}_{\gamma}$ of eigenstates of both
$J_p\cdot J_p$ as well as $J_p^3$ for all $p\in \gamma\cap H$ with
eigenvalues $\hbar^2 j_p(j_p+1)$ and $\hbar m_p$ respectively. These
states are spin network states, here denoted $|\{j_p,m_p\}_{\va
1}^{\va n}; {\van \cdots} \rangle$, where $j_p$ and $m_p$ are the
spins and magnetic numbers labeling the $n$ edges puncturing the horizon
at points $x_p$ (other labels are left implicit). They are also
eigenstates of the horizon area operator $\hat a_{\va H}$
\[ \hat a_{\va H}|\{j_p,m_p\}_{\va 1}^{\va
n}; {\van \cdots} \rangle=8\pi\beta \ell_p^2 \,
\sum_{p=1}^{n}\sqrt{j_p(j_p+1)} |\{j_p,m_p\}_{\va 1}^{\va
n}; {\van \cdots} \rangle. \]

Now substituting the expression (\ref{gammasigma})
 into the quantum version of (\ref{tress}) we get
\begin{equation}\label{seven}
-\frac{a_{\va H}}{\pi (1-\beta^2)}\epsilon^{ab}\hat{F}^i_{ab} = 16 \pi G
\beta \sum_{p \in \gamma\cap H} \delta(x,x_p) \hat{J}^i(p)
\end{equation}
As we will show now, the previous equation tells us that the surface Hilbert space, $\sH^{{\va
H}}_{\gamma\cap H}$ that we are looking for is precisely the one
corresponding to (the well studied) CS theory in the presence of {
particles}.  More precisely, consider the $SU(2)$ Chern-Simons action
\begin{eqnarray}\nonumber \label{ChernSimonssaction}
S_{\va CS}[{A}] \; = \; \frac{-a_{\va H}}{32\pi^2 G \beta
(1-\beta^2)}\int_{\Delta}{ A_i}\wedge d { A^i}  + \frac{1}{3}{
A_i}\wedge [{ A},{ A}]^i,
\end{eqnarray}
to which we couple a collection of particles by adding the following
source term:
\begin{eqnarray}\label{naivecoupling}
S_{\va \rm int}[{A},\Lambda_1{\van \cdots}\Lambda_{n}] \; =\sum_{p=1}^{n} \lambda_p
\int_{c_p}{\rm tr}[ \tau_3 (\Lambda_p^{-1}{d\Lambda_p} +
\Lambda_p^{-1} { A} \Lambda_p)], \nonumber
\end{eqnarray}
where $c_p\subset \Delta$ are the { particle} world-lines,
$\lambda_p$ coupling constants, and $\Lambda_p\in SU(2)$ are group
valued d.o.f. of the { particles}. { The particle d.o.f. being added
will turn out to correspond precisely to the d.o.f. associated to
the bulk $\hat{J}(p)^i$ appearing in (\ref{gammasigma}). The horizon
and bulk will then be coupled by identifying these d.o.f. The gauge
symmetries of the full action are
\begin{eqnarray}\label{uno}
{A} \to  gAg^{-1}+gdg^{-1}\!\!, \ \ \ \Lambda_p \to g(x_p)
\Lambda_p,\ea where $g \in C^{\infty}({\Delta},SU(2))$, and \be
\Lambda_p\to \Lambda_p \exp({\phi \tau^3})\label{dos} \ee where
$\phi \in C^{\infty}(c_p,[0,2\pi])$. 
%%%%%%%%%%%%%%%%%%%%

In order to perform the canonical analysis we assume that $\Lambda_p(r)=\exp
(-r_p^{\alpha}\tau_{\alpha})$ ($\alpha=1,2,3$). Under the left action of the group
we have
\be \exp(-\kappa^{e}\tau_{e})\Lambda_p(r)=\Lambda_p(f(r,\kappa))\ee
for a function $f(r,\kappa)$ whose explicit form will not play any role in
what follows. The infinitesimal version of the previous action is
\be \label{util} -\tau_{e} \Lambda_p(r)=\frac{\partial \Lambda_p}{\partial
  r^{\alpha}}\frac{\partial f^{\alpha}}{\partial \kappa^{e}}\ee
If we define the (spin) momentum $S^i_p$ as
\be\label{spinmom} S_p^i=-\pi^r_{\alpha} \frac{\partial
  f^{\alpha}}{\partial \kappa^{i}}, \ee
where $\pi^r_{\alpha}$ are the conjugate momenta of $r^{\alpha}$ then it is
easy to recover the following simple Poisson brackets
\ba\nonumber && \{S_p^{\alpha}, \Lambda_{p^{\prime}}\}=-\tau^{\alpha} \Lambda_p \ \delta_{pp^{\prime}}\\
&& \{S_p^{\alpha},S_{p^{\prime}}^{\beta}\}=\epsilon^{\alpha \beta}_{\ \ \gamma} S_p^{\gamma}\ \delta_{pp^{\prime}},\label{PB}
\ea
where the last equation follows from the Jacobi identity. Explicit
computation shows that $S_p^i=\lambda_p {\rm Tr}[\tau^i\Lambda_p
  \tau_3\Lambda_p^{-1}]$. Therefore;  we have three primary constraints per particle
\begin{eqnarray}\label{yyyy}
\Psi^i(S_{p},\Lambda_p) \; \equiv \; S_p^i - \lambda_p {\rm Tr}[\tau^i\Lambda_p
  \tau_3\Lambda_p^{-1}] \; \approx \; 0 \;, 
\end{eqnarray}
The primary
Hamiltonian is simply given by
\[ H(\{S_{p}\},\{\Lambda_p\})=\sum_p\eta^p_i\ \Psi^{i}(S_{p},\Lambda_p)\]
the requirement that the constraints be preserved by the time evolution reads
\ba
&& \{\Psi_i (S_p,\Lambda_p), H\}\approx -\epsilon_{ij}^{\ \, k}{\rm Tr}[\tau_i\Lambda_p
  \tau_3\Lambda_p^{-1}] \eta_p^j
\label{algebb}\ea
and the constraint algebra is
\ba
&& \nonumber\{\Psi_i(S_p,\Lambda_p), \Psi_j(S_{p^{\prime}},\Lambda_{p^{\prime}})\}\approx\epsilon_{ij}^{\ \,
k}(\Psi_k(S_p,\Lambda_p)-\lambda_p  {\rm Tr}[\tau_i\Lambda_p
  \tau_3\Lambda_p^{-1}] )\ \delta_{pp^{\prime}}. \ea If we write $\eta^p=\eta^p_{\bot} +\eta^p_{\va ||}$, where $\eta^p_{\bot}$ is the component normal to  $\Lambda_p
\tau_3\Lambda_p^{-1}$ while $\eta^p_{\va ||}$  is the parallel one, equations (\ref{algebb}) completely fix  the 
Lagrange multipliers $\eta^p_{\bot}$.  This means that, per particle, two (out of three) constraints are second class.  
The fact that $\eta^p_{\va ||}$ remains unfixed by the equations of motion implies the presence of first class contraints which are in fact given by
\begin{eqnarray}\label{primaryconstraints}
S_p \cdot S_p - \lambda_p^2 \; \approx \; 0 \;.
\end{eqnarray}
This constraint generates rotations $\Lambda\rightarrow \exp{\phi \tau_3} \Lambda$ which conserve the
quantity ${\rm Tr}[\tau_i\Lambda
\tau_3\Lambda^{-1}]$.
Now, the fact that there are secons class constraints implies that in order to quantize the theory one has to either work with Dirac brackets, solve the constraints classically 
before quantizing, or parametrize the phase space in terms of Dirac observables. In this case the third option turns out to be immediate. The reason is that the $S_p$ turn out to 
be Dirac observables of the particle system as far as the constraints $(\ref{yyyy})$ is concerned, namely
\be
\{S^i_p, \Psi^j(S_{p^{\prime}},\Lambda_{p^{\prime}})\}=\epsilon^{ijk}\Psi_k(S_{p},\Lambda_{p}) \ \delta_{pp^{\prime}}\approx 0 .
\ee
This implies that the Poisson bracket relations (\ref{PB}) remain unchanged when one replaces the brackets $\{,\}$ by Dirac brackets $\{,\}_{\va D}$.
Due to this fact and for notational simplicity we shall keep using the standard Poisson bracket notation. 

%%%%%%%%%%%%%%%%%%%%%

In summary, the phase space of each
particle is $T^{\star}(SU(2))$,  where the momenta conjugate to
$\Lambda_p$ are given by $S^i_p$
%%%%%%%%%%%%%%%%%%%%%%%%%%%%%%%%%%%%%%%%%%%%%%%%%%%%%%%%%%%%%%%%%
%\footnote{We assume that $\Lambda(r)=\exp
%(-r^{\alpha}\tau_{\alpha})$. Under the left action of the group we have
%$\exp(-\kappa^{e}\tau_{e})\Lambda(r)=\Lambda(f(r,\kappa))$ for a
%function $f(r,\kappa)$ whose explicit form will not be play any role
%in what follows. The infinitesimal version of the previous action is
%$-\tau_{e} \Lambda(r)={\partial \Lambda}/{\partial
%r^{\alpha}}\frac{\partial f^{\alpha}}{\partial \kappa^{e}}$. The
%momenta $E_i$ are defined as $E_i=-\pi^r_{\alpha} {\partial
%f^{\alpha}}/{\partial \kappa^{i}}$ where $\pi^r_{\alpha}$ are the
%conjugate momenta of $r^{\alpha}$.},
%%%%%%%%%%%%%%%%%%%%%%%%%%%%%%%%%%%%%%%%%%%%%%%%%%%%%%%%%%%%%%%%%%%
satisfying the Poisson bracket relations \ba \{S^i_{p},
\Lambda_{p^{\prime}}\}=-\tau^i \Lambda_p\ \delta_{pp^{\prime}}\ \ \ {\rm and}\ \ \
\{S_p^{i},S_{p^{\prime}}^j\}=\epsilon^{ij}_{\ \ k} S^{k}_p \ \delta_{pp^{\prime}}. \label{popo} \ea
Explicit computation shows that $S^i_p +\lambda_p \ {\rm
tr}[\tau^i\Lambda_p \tau^3\Lambda_p^{-1}]=0$ are primary constraints (two
of which are second class). In the Hamiltonian framework we use
$\Delta=H\times \R$, and the symmetries (\ref{uno}) and (\ref{dos})
arise from (and imply) the following set of first class constraints on
$H$:
\begin{eqnarray}\label{puncty}
  -\frac{a_{\va
H}}{\pi (1-\beta^2)}\epsilon^{ab}{F}_{ab}(x) &=&  {16\pi G
\beta}\sum_{p=1}^{n} \delta(x,x_p) S_p,\\ \label{primaryconstraints}
S_p\cdot S_p -\lambda_p^2 &=& 0.
\end{eqnarray}
The first constraint tells us that the level of the Chern-Simons theory is \footnote{ \label{footynoty}If we use the connection $\bar A^i$ introduced in the remark below (\ref{BIsylstr}) then the level takes the form $ k\equiv
a_{\va H}/(4\pi \ell_p^2\beta (1-\bar\beta^2))$.}\be
 k\equiv
a_{\va H}/(4\pi \ell_p^2\beta (1-\beta^2)), \label{level} \ee and that the curvature of the Chern-Simons connection vanishes
everywhere on $H$ except at the position of the defects where
we find conical singularities of strength proportional
to the defects' momenta.

The theory is topological  which means in our case that non trivial
d.o.f. are only present at punctures.
%
%At a puncture the
%(unconstrained) phase space is parametrized by 6 components of
%$A_a^i$ in addition to the 3 components of the particle
%configuration variable $\Lambda_p$ plus 3 conjugate momenta. We have
%4 first class and 2 second class constraints at that point (3
%curvature constraints (\ref{puncty}) and 1 constraint
%(\ref{primaryconstraints}) plus the 2 remaining second class
%constraints among the primary constraints defining the momenta
%$S_p^i$ (right below eq. (\ref{popo})).
%
%A naive counting yields a single d.o.f. per puncture.
Note that due to (\ref{popo}) and (\ref{primaryconstraints}) the
$\lambda_p$ are quantized according to $\lambda_p=\sqrt{s_p(s_p+1)}$
where $s_p$ is a half integer labelling  a unitary irreducible
representation of $SU(2)$.

From now on we denote $\sH^{\va CS}({s_1 {\va \cdots} s_{n}})$ the
Hilbert space of the CS theory associated with a fixed choice of
spins $s_p$ at each puncture $p\in \gamma\cap H$. { This will be a
proper subspace of the `kinematical' Hilbert space $\sH^{\va
CS}_{kin}({s_1 {\va \cdots} s_{n}}):= s_1 \otimes \cdots \otimes
s_n$. In particular} there is an important global constraint that
follows from (\ref{puncty}) { and the fact that} the holonomy around
a contractible loop that goes around all particles { is} trivial. {
It} implies \be \sH^{\va CS}(s_1 {\van \cdots } s_n) \subset {\rm
Inv}(s_1\otimes{\van \cdots}\otimes s_{n}).\label{hihi}\ee Moreover,
the above containment becomes an equality in the limit $k\equiv
a_{\va H}/(4\pi \ell_p^2\beta (1-\beta^2))\to \infty$ \cite{witten},
i.e. in the large BH limit.  In that limit we see that the
constraint (\ref{puncty}) has the simple effect of projecting the
particle kinematical states in $s_1\otimes{\van \cdots}\otimes
s_{n}$ into the $SU(2)$ singlet.
%
%\be
%|\iota \rangle=\sum_{\{m_p\}} \iota^{s_1\cdots s_n}_{m_1\cdots m_n} |m_1\cdots m_p\rangle
%\ee

{

To make contact with the bulk theory, we first note that the bulk Hilbert space $\sH^B_\gamma$
can be written
\begin{equation}
\sH_\gamma^B = \underset{\{j_p\}_{p \in \gamma \cap H}}{\bigoplus} \sH_{\{j_p\}}
\otimes \left(\underset{p \in \gamma \cap H}{\otimes}{j_p}\right)
\end{equation}
for certain spaces $\sH_{\{j_p\}}$, and where, for each $p$, the generators $\hat{J}(p)^i$
appearing in (\ref{gammasigma}) act on the representation space ${j_p}$.
%
% in each summand.
%
%As mentioned earlier, the $\hat{J}(p)^i$ and $\hat{S}(p)^i$ are actually the same:
If we now identify $\left(\otimes_p {j_p}\right)$ with the
`kinematical' Chern-Simons Hilbert space $\sH^{\va CS}_{kin}({j_1
{\va \cdots} j_{n}})$, the $J^i(p)$ operators in (\ref{seven}) are
identified with the $S^{i}(p)$ of (\ref{puncty}). The constraints of
the CS theory then restrict $\sH^{CS}_{kin}$ to $\sH^{CS}$ yielding
\begin{equation} \sH_\gamma = \underset{\{j_p\}_{p \in \gamma \cap
H}}{\bigoplus} \sH_{\{j_p\}} \otimes \sH^{\va CS}({j_1 {\va \cdots}
j_{n}}),
\end{equation}
as the full kinematical Hilbert space for \mbox{fixed $\gamma$}.

}

So far we have dealt with a fixed graph. The Hilbert space satisfying the quantum version of
(\ref{tress}) is obtained as the projective limit of the spaces
$\sH_\gamma$.
%
%\footnote{To define the
%projective limit, one needs to define a family of consistent embeddings
%$\iota_{\gamma' \gamma} : \sH_{\gamma}^{kin} \rightarrow \sH_{\gamma'}^{kin}$
%whenever each edge of $\gamma$ can be obtained from edges of $\gamma'$ by
%composition and reversal of orientation.  One can use the usual embeddings
%appropriate for LQG --- namely pull-back via the projection $p_{\gamma'\gamma}$
%defined, e.g., in \cite{aldiffproj, alrev}; one can check that these embeddings
%preserve the $SU(2)$-invariance of the states, as well as the horizon invariance condition above,
%and therefore are also well-defined on the spaces $\sH_\gamma^{kin}$, and satisfy the
%necessary consistency conditions.}
%Finally, in order to obtain the physical Hilbert space it remains
%only to impose the diffeomorphism, and Hamiltonian constraints.
The Gauss and diffeomorphism constraints are imposed in the same way as in \cite{ack, bhe1}.
%The resulting Hilbert space $\sH^{Diff}$ is spanned by states which we may
%represent as $|n; \{j_p,m_p\}_{\va 1}^{\va n}; {\van \cdots} \rangle$.  Here $n$ is the
%number of punctures at the horizon, and $\{j_p,m_p\}_{\va 1}^{\va n}$
%is an ordered set of pairs $\{j_p,m_p\}$, one for each puncture; these are the data from the
%horizon Hilbert space that remain.  $\dots$ represent the rest of the data specifying the
%bulk state (which will not be relevant for this discussion).
The IH boundary condition implies that lapse must be zero at the
horizon so that the Hamiltonian constraint is only imposed in the
bulk.
%Because
%there is more horizon data in this case, this assumption is slightly
%stronger than in \cite{ack}. }

\subsection{State counting}

The entropy of the IH is computed by the formula $S={\rm tr}(\rho_{\va IH}\log\rho_{\va IH})$ where the density matrix
$\rho_{\va IH}$ is obtained by tracing over the bulk d.o.f., while restricting to horizon states that are compatible
with the macroscopic area parameter $a_{\va H}$. Assuming that there exist at least
one solution of the bulk constraints for every state in the CS theory, the entropy is given by
$S=\log(\sN)$ where $\sN$ is the number of  horizon states compatible with the given
macroscopic horizon area $a_{\va H}$. After a moment of reflection one sees that
\be {\sN}=\sum_{n;(j)_{\va 1}^{\va
n}}
%{\rm Sym}[n;\{j\}_{\va 1}^{\va n}]\
{\rm dim}[{\sH}^{\va CS}(j_1 {\van \cdots} j_n)], \ee where the
labels $j_1\cdots j_p$ of the punctures are constrained by the condition
\be \label{conki}a_{\va H}-\epsilon \le 8 \pi\beta
\ell_p^2 \, \sum_{p=1}^{n}\sqrt{j_p(j_p+1)}\le a_{\va H}+\epsilon . \ee
%
%The symmetry factor---given by \be {\rm
%Sym}[n;\{j\}_{\va 1}^{\va n}]\equiv \frac{n!}{\prod_j n_j!},
%\ee where $n_j$ denotes the number of punctures labelled by the spin
%$j$, { so that} $n=\sum_j n_j$---comes from the correct
%implementation of diffeomorphism invariance on the horizon
%\cite{bhe}.
Similar  formulae, with a different $k$ value, were first used in \cite{majundar}.

Notice that due to (\ref{conki}) we can compute the entropy for
$a_{\va H}>>\beta\ell_p^2$ (not necessarily infinite).  The reason
is that the representation theory of $U_q(SU(2))$---describing
${\sH}_{\va CS}$ for finite $k$---implies \be{\rm dim}[{\sH}_{\va
CS}(j_1 {\van \cdots} j_n)]={\rm dim}[{\rm Inv}(\otimes_p j_p)],\ee
as long as all the $j_p$ as well as the interwining internal spins
are less than $k/2=a_{\va H}/(8\pi\beta(1-\beta^2)\ell_p^2)$. But for
Immirzi parameter in the range $|\gamma|\le\sqrt{2}$ this is
precisely granted by (\ref{conki}) according to the Lemma below. All this simplifies the entropy
formula considerably. The previous dimension corresponds to the
number of independent states one has if one models the black hole by
a single $SU(2)$ intertwiner!

\noindent {\bf Lemma 5: } The Hilbert spaces  ${\sH}^{\va CS}(j_1 {\van \cdots} j_n)$ of Chern-Simons theory with level $k$ selected by the  restriction 
\be\label{condi}\sum_{p=1}^{n}\sqrt{j_p(j_p+1)}\le \frac{k}{2}\ee are isomorphic to  ${\rm Inv}[(j_1 {\van \cdots} j_n)]$.

\noindent{\em Proof:} 
The Chern-Simons Hilbert space $\mathscr{H}^{CS}(j_1\cdots j_n)$
will be isomorphic to ${\rm Inv}[(j_1 \cdots j_n)]$ if for instance
all element of a given basis (of intertwiners) of ${\rm Inv}[(j_1 {\van \cdots} j_n)]$  if  (voir for instance \cite{babaez}) \be \label{primo} j_p\le \frac{k}{2}\ \ \ \forall \ \ p=1,\cdots,n\ee and  \be \iota_a\le \frac{k}{2}\ \ \ \forall \ \ a=1,\cdots, n-3. \label{seco}\ee  
Equation (\ref{primo}) is immediately implied by (\ref{condi}) as the latter implies \be\label{condimento} \sum_{p} j_p \le \frac{k}{2}. \ee The condition (\ref{seco}) requires a more precise analysis.
Notice the fact that, being intertwining spins, the $\iota_a$ satisfy the following set of nested restrictions which imply the result:
\ba
&& 
\nonumber 0 \le \iota_1\le {\rm Min}[j_1+j_2,j_3+\iota_2] \le j_1+j_2\le \frac{k}{2}\\ && \nonumber  0 \le \iota_2 \le {\rm Min}[j_3+\iota_{1},j_{4}+\iota_{3}]\le j_1+j_2+j_3\le \frac{k}{2}\\ \nonumber && \ \ \ \ \ \ \ \ \cdots\\  
&& \nonumber 0 \le \iota_{n-4} \le {\rm Min}[j_{n-3}+\iota_{n-5}, j_{n-2}+\iota_{n-3}]\le \sum_{p=1}^{n-3} j_p \le \frac{k}{2}\\ && \nonumber  0 \le \iota_{n-3} \le {\rm Min}[j_{n-2}+\iota_{n-4},j_n+j_{n-1}]\le \sum_{p=1}^{n-2} j_p \le \frac{k}{2},
\ea 
where in each line we have used (\ref{condimento}) $\square$.

{ {\em Remark:} An interesting point can be made here as a further developement of the remark below equation (\ref{BIsylstr}). Notice that if we had worked with the connection $\bar A^i=\Gamma^i$ as our boundary field degree of freedom---corresponding to the choice $\bar \beta=0$ in the notation of the remark below (\ref{BIsylstr})---then the boundary Chern-Simons level would be $k=a_{\va H}/(4\pi\ell_p^2 \beta)$ (see Footnote \ref{footynoty}). This implies that the condition (\ref{primo}) imposed on representations labelling the punctures  would take the simple form \be j_{p}\le j_{\va max}\equiv \frac{a_{\va H}}{8\pi \ell_p^2\beta}, \ee
or equivalently
 \be
  j\le j_{\va max} \ \ \ \ s.t.\ \ \ \ a^{\va (1)}_{\va max} = 8\pi \ell_p^2\beta \sqrt{j_{\va max}(j_{\va max}+1)} \approx 8\pi \ell_p^2\beta j_{max}=a_{\va H}
 \ee
where $a^{\va (1)}_{\va max}$ is the maximum single-puncture eigenvalue allowed. 
Our effective treatment depends on a classical input: the macroscopic area. 
 One would perhaps hope that this effective treatment would only allow for states where the microscopic area is {\em close} to $a_{\va H}$, unfortunately such a strong requirement is not satisfied as the allowed eigenvalues can be very far away from $a_{\va H}$. However,  the effective theory at least  forbids  quantum states where individual area quanta are larger than $a_{\va H}$. This is a nice interplay between the classical input and the associated  effective quantum description. Of course this interplay is still qualitatively valid for the case in which one works with the Ashtekar-Barbero connection on the boundary (i.e., $\beta=\bar \beta$).}

\section{Conclusion} \label{conclu}

We have shown that the spherically symmetric isolated horizon (or Type I isoated horizon) can be
described as a dynamical system by a pre-symplectic form $\Omega_M$ that, when written in the
(connection) variables suitable for quantization, acquires a horizon
contribution corresponding to an SU(2) Chern-Simons theory. There are different ways to 
prove this important statement. In \cite{nous}
we first observed that $SU(2)$ gauge transformations and diffeomorphism preserving $H$
are not broken by the IH boundary condition. Moreover,  infinitesimal diffeomorphisms 
tangent to $H$ and $SU(2)$ local transformations continue to be degenerate directions of $\Omega_M$ on shell. This by
itself is then sufficient for deriving the boundary term that arises
when writing the symplectic structure in terms of Ashtekar-Barbero
connection variables. Here we have reviewed this construction in Section \ref{GSS}.
{ A result that was not explicitly presented in \cite{nous} is the precise form of the constraint algebra found in Subsection \ref{CCA}.
There we see in a precise way how the canonical gauge symmetry structure of our system is precisely that of an $SU(2)$ Chern-Simons theory: in particular, at the boundary, 
infinitesimal diffeomorphisms, preserving $H$, form a subalgebra of $SU(2)$ gauge algebra, as in the topological theory.}

A different, more direct approach is based on a subtle
fact about the canonical transformation that takes us from the
Palatini $(\Sigma^i_{ab}, K^i_a)$ phase space parametrization
to the Ashtekar-Barbero $(\Sigma^i_{ab}, A^i_a)$ connection
formulation, in the presence
of an internal boundary. In the case of Type I isolated horizons, the term to be added to the symplectic potential producing the above transformation gives rise to a boundary contribution that
eventually leads to a boundary Chern-Simons term in the presymplectic structure. This is the content of 
Section \ref{barbero}. The boundary Chern-Simons term appears due to the use of connection variables
which in turn are the ones in terms of which the quantization program of loop quantum gravity is applicable.

Finally, at a fundamental level, what actually fixes the surface term in the symplectic structure is the requirement 
that it be conserved in time. The above mentioned proofs show that the various expressions for the symplectic structure
using different variables are in fact one and the same symplectic structure. That this symplectic structure is preserved in time was proven in Section \ref{CSS}.

{ There is a certain freedom in the choice of boundary variables leading to different parametrizations of the boundary 
degrees of freedom. The most direct description would appear, at first sight, to be  the one defined simply in terms of the triad field (pulled back on H)  along the lines exhibited in Section \ref{sidy}.
Such parametrization is however less preferable from the point of view of quantization as one is confronted to the background independent quantization of form fields for which the usual techniques
are not directly applicable. In contrast, the parametrization of the boundary degrees of freedom in terms of a connection directly leads to a description in terms of  $SU(2)$ Chern-Simons theory 
which, being a well studied topological field theory, drastically simplifies the problem of quantization. However, such description comes with the freedom of the introduction of an extra dimensionless parameter
$\bar \beta$ (as pointed out in the Remark below equation (\ref{BIsylstr})). Such appearance of extra parameters  is very much related to what happens in the general context 
of the canonical formulation of gravity in terms of connections (see Appendix in \cite{rezendeyo}). Therefore, this observation is by no means a new feature proper of IHs.
The existence of this extra parameter has a direct influence on the value of the Chern-Simons level; however, the value of the entropy is independent of this extra parameter \cite{future2}.
}

Note that no d.o.f. is available at the horizon in the classical
theory as the IH boundary condition completely fixes the geometry at
$\Delta$ (the IH condition allows a single (characteristic) initial
data once $a_{\va H}$ is fixed (see fig. \ref{figui})).
Nevertheless, non trivial d.o.f. arise as {\em would be gauge}
d.o.f. upon quantization.  These are described by SU(2) Chern-Simons theory
coupled to (an arbitrary number of) defects through
a dimensionless parameter 
proportional to $4\pi (1-\beta^2)a_j/a_{\va H}$, where $a_j=8\pi \ell_p^2 \sqrt{j({j+1})}$ is the basic quantum of area carried by the defect. 
These {\em would be gauge} excitations
are entirely responsible for the entropy in this approach \footnote{More insight on the nature of these degrees of freedom
could be gained by studying simpler models. In \cite{liu} a theory with no local degrees of freedom has been introduced.
The attractive feature of this model is that it admits an (unconstrained) phase parametrization in terms of the 
same field content as gravity. Moreover, one can argue that it contains the minimal structure to serve as a toy model 
to study some generic features of the Type I isolated horizon quantization.}.

We obtain a remarkably simple formula for the horizon entropy: the
number of states of the horizon is simply given in terms of the
(well studied) dimension of the Hilbert spaces of Chern-Simons
theory with punctures labeled by spins. In the large $a_{\va H}$
limit the latter is simply equal to the dimension of the singlet
component in the tensor product of the representations carried by
punctures. In this limit the black hole density matrix $\rho_{\va
IH}$  is the identity on ${\rm Inv}(\otimes_p j_p)$ for admissible
$j_p$. Similar counting formulae have been proposed in the
literature \cite{models, majundar}. Our derivation from first principles clarifies these
previous proposals.

Remarkably, the counting of states necessary to compute the entropy
of the above Type I isolated horizons can be exactly done \cite{barberos} using the novel counting techniques introduced in \cite{barba}. It turns
out to be $S_{\va BH}={\beta_0} a_{\va H}/({4 \beta \ell^2_p})$,  where
$\beta_0=0.274067 ...$.  However, the
subleading corrections turn out to have the form $\Delta S=-\frac{3}{2}
\log a_{\va H}$ (instead of the $\Delta S=-\frac{1}{2} \log a_{\va H}$ that
follows the classic treatment \cite{bhe1,amit}) matching other approaches
\cite{majundar}. This is due to the full $SU(2)$ nature of the IH
quantum constraints imposed here. We must mention that the proposal of Majumdar et al. \cite{majundar} is 
most closely related to our result. Their intuition was particularly
insightful as it yielded a universal form of logarithmic corrections
in agreement with those found in different quantum gravity 
formulations \cite{carlip-log}.  
Our work clarifies the relevance of their proposal.

{  We have concentrated in this work on Type I isolated horizons. The natural question that follows from this analysis is whether we can 
generalize the $SU(2)$ invariant treatment in order to include distortion. The classical formulation and quantization of Type II 
isolated horizons in the $U(1)$ (gauge fixed) treatment has been studied in  \cite{jon}. Work in progress \cite{future1} shows that, in the $SU(2)$ invariant formulation, 
it is possible to include  distortion in a simple way as long as the isolated horizon is non rotating (i.e., when ${\rm Im}[\Psi_2]=0$). The rotating case is more subtle but we believe that there are no
insurmountable obstacles to its $SU(2)$ invariant treatment (this will be studied elsewhere).}

\section{Acknowledgements}

We thank A. Ashtekar, C. Beetle, E. Bianchi, K. Krasnov, M. Montesinos, M. Reisenberger, and C. Rovelli for discussions.
This work was supported in part by the Agence Nationale de la Recherche; grant
ANR-06-BLAN-0050. J.E. was supported by NSF grant OISE 0601844
and the {\em Alexander von Humboldt Foundation} of Germany,
and thanks Florida Atlantic University
for hospitality during his visit there. A.P. was supported by {\em l'Institut Universitaire de France}. D.P. was supported by {\em Marie Curie} EU-NCG network.

\begin{appendix}

\section{Type I Isolated Horizons: Horizon geometry from the Reissner-Nordstrom family}\label{direct}

The spherically symmetric isolated horizons or Type I  isolated horizons are easy
to visualise in terms of the characteristic formulation of general
relativity with initial data given on null surfaces \cite{lewa}. 
This observation is very useful if one is looking for a concrete visualisation of the horizon geometry and 
properties of the matter fields at the horizon. In this appendix we chose to derive the main properties of Type I 
isolated horizons by studying their geometry in the context of Einstein-Maxwell theory (which is general enough for
the most relevant applications of the formalism). An additional motivation for the explicit approach presented here 
is its complementarity with more abstract discussions available in the literature \cite{ack, better, ih_prl}.
In the context of Einstein-Maxwell theory, 
spacetimes with a
Type I IH are solutions to Einstein-Maxwell equations where Reissner-Nordstrom 
horizon data are given on a null surface $\Delta=S^2\times \R$ and suitable free radiation is given
at the transversal null surface for both geometric as well as electromagnetic degrees of freedom. This allows to derive the main
equations of IH directly from the Reissner-Nordstrom geometry as far as we are careful enough only to use
the information that is intrinsic to the IH geometry.

\subsection{The Reissner-Nordstrom solution in Kruskal-like coordinates}
%%%%%%%%%%%%%%%%%%%%%%%%%%%%%%%%%%%%%%%%%%%%%%%%%%%%%%%%%%%%%%%%%%%%%%%%%%%%%%
The Reissner-Nordstrom metric can be written in Kruskal-like coordinates \cite{chandra} as
\be ds^2=\Omega^2(x,t) (-dt^2+dx^2)+
r^2(d\theta^2+\sin(\theta)d\phi^2)\ee where 
 \be \Omega(x,t)=\frac{(r-r_{-})^{\frac{1+b}{2}}e^{-a r}}{ar},\ee
with $a=(r_+-r_-)/(2r_+^2)$, $b=r_-^2/r_+^2$, and the function $r(x,t)$ is
determined by the following implicit equation: \be
F(r)=x^2-t^2;\ \ \ {\rm with }\ \ \ F(r)=\frac{(r-r_+)e^{2ar}}{(r-r_-)^{b}} .\ee 
The previous Kruskal-like coordinates are valid for the external region $r\ge r_+$. The metric is smooth at the horizon $r=r_+$ which in the new coordinates corresponds to the null surface  $x=t$.
An
important identity is: \be dr|_{\Delta}=\frac{2x}{F^{\prime}}\ (dx-dt), \ee
where $|_{\Delta}$ denotes that the equality holds at the horizon
$\Delta$ for which $x=t$. Here we are interested in the first order
formalism. Thus we are interested in an associated tetrad $e_\mu^I$
with $I=0,1,2,3$. It is immediate to verify that a possible such
tetrad is given by \ba \begin{array}{lll} e^0=\Omega(x,t)
dt \\ 
e^1=\Omega(x,t) dx\end{array}\ \ \ \  \begin{array}{lll} e^2=r d\theta \\  e^3=r\sin(\theta)
d\phi\end{array} \ea We now want to compute the components of
the spin connection $\omega_a^{IJ}$ at the horizon. Therefore, we
will use Cartan's first structure equations $de+\omega\wedge e=0$
at $\Delta$. The solution is (all details are given in Section \ref{ce})  
 \ba
\begin{array}{lll} \omega^{01}\eh  \frac{2x\Omega^{\prime}}{ F^{\prime}\Omega} \ (dt-dx)
\\ \omega^{02}\eh -\frac{2x}{F^{\prime}\Omega} \ d\theta\\ 
 \omega^{03}\eh  -\frac{2x}{F^{\prime}\Omega} \sin(\theta) \ d\phi \end{array}\ \ \ \ \begin{array}{lll}
\omega^{12}\eh  -\frac{2x}{F^{\prime}\Omega} \ d\theta \\ 
\omega^{13}\eh  -\frac{2x}{F^{\prime}\Omega} \sin(\theta) \ d\phi\\
\omega^{23}\eh- \cos(\theta) \ d\phi .\end{array} \label{resulti}\ea 
At this stage we consider a
Lorentz transformation of the form \be
\Lambda^{I}_J=\left[\begin{array}{ccc}  c\ \  s\ \ 0\ \ 0\\  s\ \  c\ \  0\ \  0 \\  0\ \ 0\ \ 1\ \ 0 \\
 0\ \ 0\ \ 0\ \ 1
\end{array}\right],\label{lt}
\ee where $c=\cosh(\alpha(x))$ and $s=\sinh(\alpha(x))$. It is
immediate to see that under such transformation the connection above
transforms to \ba \begin{array}{lll}\spb{\tilde\omega^{01}}=
-\alpha^{\prime}(x)\ dx\\ \spb{\tilde \omega^{02}}=
-\lambda(x) \ d\theta\\\spb{\tilde \omega^{03}}=
-\lambda(x) \sin(\theta) \ d\phi  \end{array}\ \ \ \ \begin{array}{lll}  \spb{\tilde
\omega^{12}}= -\lambda(x) \ d\theta \\ \spb{\tilde
\omega^{13}}= -\lambda(x) \sin(\theta) \ d\phi\\ \spb{\tilde
\omega^{23}}= -cos(\theta) \ d\phi \end{array}\label{exacte}\ea where the arrows
below the components denote the pull-back of the one forms to $\Delta$
, and $\lambda(x)=\frac{2x}{F^{\prime}\Omega} \exp(\alpha(x))$.  We can obviously chose this Lorentz transformation in
order for $\lambda(x)=\lambda_0$ with $\lambda_0$ an arbitrary
constant. We have\be\label{l0} \lambda_0= \frac{2x}{F^{\prime}\Omega}
\exp(\alpha_0(x)) \ee This can be made compatible with the time gauge
by changing the spacetime foliation just at the intersection with the
horizon $\Delta$ so that $\tilde e^0=(\Lambda\cdot e)^0$ is the new
normal\footnote{ Recently, a similar analysis as the one presented here---and also in \cite{beigin}---has been done \cite{maju}. In that reference the authors derive a result which is compatible with the above equations in the {\em singular} vanishing extrinsic curvature slicing $\lambda_0=0$. Such (null) slicing is however inconsistent with the canonical formulation that is necessary for the LQG quantization of the bulk degrees of freedom.  }. Now we are ready to write the quantities we were looking
for\ba \begin{array}{lll}\sdpb{K^1}= 0 \\ \sdpb{K^2}= -\lambda_0 \
d\theta\\ \sdpb{K^3}= -\lambda_0 \sin(\theta) \ d\phi \end{array}\ \ \
\ \begin{array} {lll} \sdpb{\Gamma}^3= \lambda_0 \ d\theta \\
\sdpb{\Gamma^2} = -\lambda_0 \sin(\theta) \ d\phi\\ \sdpb{\Gamma^1}=
\cos(\theta) \ d\phi\end{array} \label{k1}\ea where $\Gamma^i=-
\frac{1}{2}\epsilon^{ijk} \omega_{jk}$ and $K^i=\omega^{0i}$. The self dual connection
$A^{i}_{\va +}\equiv \Gamma^i+i K^i$ and the Ashtekar-Barbero
connection become \ba \begin{array}{lll} \sdpb{A_{\va +}^3}= 
\lambda_0\ (-i\ \sin(\theta)d\phi+ d\theta)\\ \sdpb{A_{\va +}^2}=
\lambda_0\ ( -\sin(\theta) \ d\phi-i\ d\theta)\\ \sdpb{A_{\va +}^1} =
\cos(\theta) \ d\phi \end{array}\ \ \ \ \begin{array}{lll} \sdpb{A_{\va
\beta}^3}= \lambda_0\ (-\beta\ \sin(\theta)d\phi+ d\theta) \\
\sdpb{A_{\va \beta}^2}= \lambda_0\ ( -\sin(\theta) \ d\phi-\beta\
d\theta) \\ \sdpb{A_{\va\beta}^1}= \cos(\theta) \ d\phi \end{array} \ea The curvature of the self-dual
and Ashtekar-Barbero connections is (when pulled back to the cross
sections $H$)  \ba \begin{array}{lll} \sdpb{F^3_{\va +}}= 0 \\  \sdpb{F^2_{\va +}}= 0\\
\sdpb{F^1_{\va +}} =
-\sin(\theta)\ d\theta\wedge d\phi 
\end{array}
\ \ \ \ \begin{array}{lll} \sdpb{F^3_{\va \beta}}= 0 \\ \sdpb{F^2_{\va
\beta}}=0 \\ \sdpb{F^1_{\va \beta}} =
-(1-\lambda^2_0[1+\beta^2])\sin(\theta)\ d\theta\wedge
d\phi\end{array} \ea Using that $a_{\va H}=4\pi r^2$ we can write the previous equations as 
\be
\sdpb{F_{\va +}^i}=-\frac{2\pi}{a_{\va H}}\sdpb{\Sigma^i}
\ee
and
\be\label{main}
\sdpb{F^i_{\va \beta}}=(1-\lambda^2_0 (1+\beta^2))\ \sdpb{F_{\va +}^i}=
-\frac{2\pi (1-\lambda^2_0 (1+\beta^2))}{a_{\va H}}\ \sdpb{\Sigma^i}.
\ee 
In the following subsection we will show that $\lambda_{0}=-1/\sqrt{2}$ defines the frame where the IH surface gravity matches the stationary black hole one. With this value of $\lambda_0$, the
previous two equations and equation (\ref{k1}) imply eqs. (\ref{unos}), (\ref{doss}), and (\ref{tress}) respectively.
For completeness we write the componets of $\Sigma^{IJ}$ \ba \begin{array}{lll} 
\spb{\Sigma^{01}}=0\\ \spb{\Sigma^{02}}=r\Omega\exp(\alpha)
\ dx\wedge d\theta \\ \spb{\Sigma^{03}}=r\Omega\exp(\alpha)
\sin(\theta) \ dx\wedge d\phi \\ 
\spb{\Sigma^{12}}=r\Omega\exp(\alpha)\ dx\wedge d\theta \\ \spb{\Sigma^{13}}=r\Omega\exp(\alpha) \sin(\theta) \ dx\wedge d\phi \\ \spb{\Sigma^{23}}=r^2 \sin(\theta)\ d\theta\wedge d\phi\end{array}\ \ \ \ \ \begin{array}{lll}
\spb{\Sigma^{3}_{\va +}}= r\Omega\exp(\alpha)\ dx\wedge d\theta
+i r\Omega\exp(\alpha) \sin(\theta) \ dx\wedge d\phi \\ \spb{\Sigma^{2}_{\va +}}=-\exp(\alpha)\Omega r \sin(\theta) \ dx\wedge
d\phi+ i r\Omega\exp(\alpha) \ dx\wedge d\theta \\ \spb{\Sigma^{1}_{\va
+}}=r^2 \sin(\theta)\ d\theta\wedge d\phi \end{array} \label{sigg}\ea
where on the right we have written the corresponding self-dual components.

\subsection{Surface gravity and the value of $\lambda_0$}

For stationary black holes, the surface gravity $\kappa_{\va H}$ is defined by the equation
\be\label{sg}
\ell^a\nabla_a\ell^b=\kappa_{\va H}\,  \ell^b
\ee
where $\ell^a$ is the Killing vector field tangent to the horizon.
For isolated horizons there is no unique notion of $\ell^a$. 
We shall define $\ell_a$ in terms of the tetrad in the usual way
with $\ell_a \equiv (e^1_a - e^0_a)/\sqrt{2}$  \footnote{The future pointing null generators of the horizon
$\ell^a$ are such that \[\ell^a\propto (\partial/\partial x)^a+ (\partial/\partial t)^a.\] This implies that $\ell_a\propto dx_a-dt_a$ from wich we get $\ell_a=(e_a^1-e_a^0)/\sqrt{2}$ and $n_a=-(e_a^1+e_a^0)/\sqrt{2}$ so that $n\cdot \ell=-1$.}.
However, this definition still allows the freedom associated to the Lorentz 
transformations (\ref{lt}) which send $\ell^a \rightarrow \exp(-\alpha(x))\ell^a$.
We can fix this freedom by demanding the surface gravity to match that
of a Reissner-Nordstrom black hole with mass $M$ and charge $Q$ for which
\be
\kappa_{\va H}=\frac{\sqrt{(M^2-Q^2)}} {2M[M+\sqrt{(M^2-Q^2)}]-Q^2}.
\ee
Indeed this choice is the one that makes the zero, and first law
of IH look just as the corresponding laws of stationary black hole mechanics.

This choice is then physically motivated. In turn this will fix the value of $\lambda_0$
in (\ref{main}). If we define $n_a\equiv-(e_a^0+e_a^1)/\sqrt{2}$ then we have that (\ref{sg})
implies
\ba
&& \nonumber \ell^an_b\nabla_a\ell^b=-\kappa_{\va H} \\
&& \nonumber -\frac{1}{2} \ell^a (e^{0}_b+e^{1}_b) \nabla_a (e^{1b}-e^{0b})=-\kappa_{\va H} \\
&& \ell^a\omega_a^{01}=\kappa_{\va H}. \label{capa}
\ea
Notice that after the Lorentz transformation (\ref{lt}) we have
%\ba
%&& \ell^a=\exp{(-a_0)} g^{ab} \Omega\frac{e^{0}_{b}-e^{1}_{b}}{\sqrt{2}}=
%-\exp{(-a_0)} g^{xx}  \frac{\Omega}{\sqrt{2}} \frac{\partial}{\partial (x+t)}=\nonumber \\ \nonumber \\
%&&\ell^a=-\exp{(-a_0)} \frac{1}{\Omega \sqrt{2}} \frac{\partial}{\partial (x+t)}
%\ea
\ba\nonumber  \kappa_{\va H} &=&\ell^a\omega_{a}^{01}=-\alpha^{\prime} \ell^a
dx_a=-\alpha^{\prime} g^{ab}\ell_a
dx_b=-\exp{(-\alpha)}\frac{\Omega}{\sqrt{2}}\alpha^{\prime}
g^{ax}(dt_a-dx_a)=\exp{(-\alpha)}\frac{\Omega}{\sqrt{2}}\alpha^{\prime}
g^{xx}\\ &=&-(\exp{(-\alpha)})^{\prime}\frac{\Omega}{\sqrt{2}}
g^{xx}=-(\exp{(-\alpha)})^{\prime}\frac{1}{\sqrt{2}\Omega}.\ea 
%\ba
%\kappa_{\va H} &=&\exp{(-a_0)} \frac{1}{\Omega \sqrt{2}} \frac{\partial}{\partial (x+t)}(a_0^{\prime} (dx)_a)\\
%&=& \exp{(-a_0)} \frac{1}{\Omega \sqrt{2}} a_0^{\prime}\nonumber \\
%&=& -[\exp{(-a_0)}]^{\prime} \frac{1}{\Omega \sqrt{2}}
%\ea
Now we can  fix $\alpha(x)=\alpha_0(x)$ so that $\kappa_{\va H}$ takes the
RN value. Recalling equation (\ref{l0}) and using the above equations, a simple calculation shows that
this happens for  \be
\lambda_0=-\frac{1}{\sqrt{2}}\label{lambda}\ee which
implies the desired result \be \mbox{$F^i_{\va
\beta}=\frac{1}{2}(1-\beta^2)\ F_{\va +}^i$}. \ee Notice that
\be\nabla_a \ell_b=\omega^{01}_a\ell_b, \ee and that (according to
(\ref{exacte})) we also have \be d\omega^{01}=0.\ee All this implies
that $\sL_{\ell} \omega^{01}=d(\ell\inter \omega^{01})+\ell\inter
d\omega^{01}= d\kappa_{\va H}=0$ as expected from 
$[\sL_{\ell},D]=0$ (general proof in Lemma 2). In other words, the $\ell$ we have chosen 
by means of fixing the boost freedom $\ell\rightarrow \exp(-\alpha(x)) \ell$
is a member of the equivalence class $[\ell]$ in 
Definition \ref{defini}.

\subsection{Solving Cartan's equation}\label{ce}

For this we first compute $de$, namely: \ba \nonumber && de^0 =
\Omega'(x,t) dr\wedge dt \eh 2\Omega' \frac{x}{F^{\prime}}\ dx\wedge
dt \\ \nonumber && de^1\eh 2\Omega' \frac{x}{F^{\prime}}\ dx\wedge dt
\\ \nonumber && de^2\eh \frac{2x}{F^{\prime}}\ (dx\wedge
d\theta-dt\wedge d\theta) \\ && de^3\eh-r\cos(\theta)\ d\theta\wedge
d\phi + \frac{2x}{F^{\prime}}\sin(\theta)\ (dx\wedge d\phi-dt\wedge
d\phi). \ea Now we are ready to explicitly write Cartan's first
structure equations. They are \ba\label{1} &&\nonumber 0 \eh2\Omega'
\frac{x}{F^{\prime}}\ dx\wedge dt +\Omega\ \omega^{01} \wedge dx+ r
\omega^{02}\wedge d\theta + r \sin(\theta) \omega^{03}\wedge d\phi \\
\nonumber
\label{2}&& 0\eh2\Omega' \frac{x}{F^{\prime}}\ dx\wedge dt + \Omega\ \omega^{01}
\wedge dt + r \omega^{12}\wedge d\theta + r \sin(\theta)
\omega^{13}\wedge d\phi \\\label{3} \nonumber && 0 \eh
\frac{2x}{F^{\prime}}\ (dx\wedge d\theta-dt\wedge d\theta)+ \Omega
\omega^{02} \wedge dt+ \Omega \ \omega^{21}\wedge dx + r \sin(\theta)
\omega^{23}\wedge d\phi \\\label{4} \nonumber && 0 \eh -r\cos(\theta)\
d\theta\wedge d\phi + \frac{2x}{F^{\prime}}\sin(\theta)\ (dx\wedge
d\phi-dt\wedge d\phi)+ \\ && \ \ \ \ \ \ \ \ \ \ \ \ \ \ \ \ \ \ \ \ \
\ \ \ \ \ \ \ \ \ \ \ \ \ \ \ \ \ \ \ \ \ \ \ \ \ \ \ \ + \Omega
\omega^{03} \wedge dt+ \Omega \ \omega^{31}\wedge dx + r \omega^{32}
\wedge d\theta .\ea Let us now study the previous equation
individually. The six components of the first, equation (\ref{1})
become \ba \nonumber && 0\eh dx\wedge dt \ (2\Omega'
\frac{x}{F^{\prime}}-\omega^{01}_t \Omega) \\ \nonumber && 0\eh
dx\wedge d\theta \ (-\Omega\omega^{01}_{\theta} +\omega^{02}_x r)\\
\nonumber && 0\eh dx\wedge d\phi \ (-\Omega\omega^{01}_{\phi}
+\omega^{03}_x r \sin(\theta))\\ && \nonumber 0\eh dt\wedge d\theta \
(r \omega^{02}_{t})\\ && 0\eh dt\wedge d\phi \ (\omega^{03}_t r
\sin(\theta)) \nonumber \\ && 0\eh d\theta \wedge d\phi \ (-r
\omega^{02}_{\phi} +\omega^{03}_{\theta} r \sin(\theta)) .\ea The six
components of the second, equation (\ref{2}), become \ba && \nonumber
0\eh dx\wedge dt \ (2\Omega' \frac{x}{F^{\prime}}+\omega^{01}_x
\Omega) \\ &&\nonumber 0\eh dx\wedge d\theta \ (\omega^{12}_x r)\\
&&\nonumber 0\eh dx\wedge d\phi \ (\omega^{13}_x r \sin(\theta))\\
\nonumber && 0\eh dt\wedge d\theta \ (-\Omega\omega^{01}_{\theta} +r
\omega^{12}_{t})\\ && \nonumber 0\eh dt\wedge d\phi \
(-\Omega\omega^{01}_{\phi} + r \sin(\theta) \omega^{13}_{t} ) \\ &&
0\eh d\theta \wedge d\phi \ (-r \omega^{12}_{\phi}
+\omega^{13}_{\theta} r \sin(\theta)) .\ea The six components of the
third, equation (\ref{3}), become \ba &&\nonumber 0\eh dx\wedge dt \
(\omega^{02}_x \Omega+\omega^{21}_t \Omega) \\ &&\nonumber 0\eh
dx\wedge d\theta \ (2 \frac{x}{F^{\prime}}-\omega^{21}_{\theta}
\Omega)\\ && \nonumber 0\eh dx\wedge d\phi \ (-\omega^{21}_{\phi}
\Omega +\omega^{23}_x r \sin(\theta))\\ &&\nonumber 0\eh dt\wedge
d\theta \ (-2 \frac{x}{F^{\prime}}-\Omega\omega^{02}_{\theta})\\ &&
\nonumber 0\eh dt\wedge d\phi \ (-\Omega\omega^{02}_{\phi} + r
\sin(\theta) \omega^{23}_{t} ) \\ && 0\eh d\theta \wedge d\phi \
(\omega^{23}_{\theta} r \sin(\theta)). \ea Finally, the six components
of the fourth, equation (\ref{4}), become \ba && \nonumber 0\eh
dx\wedge dt \ (\omega^{03}_x \Omega-\omega^{31}_t \Omega) \\ &&
\nonumber 0\eh dx\wedge d\theta \ (-\omega^{31}_{\theta} \Omega+
\omega^{32}_{x} r)\\ && \nonumber 0\eh dx\wedge d\phi \ (2
\frac{x}{F^{\prime}} \sin(\theta) -\omega^{31}_{\phi} \Omega)\\ &&
\nonumber 0\eh dt\wedge d\theta \ (-\omega^{03}_{\theta} \Omega+r
\omega^{32}_{t})\\ &&\nonumber 0\eh dt\wedge d\phi \ (-2
\frac{x}{F^{\prime}} \sin(\theta) -\Omega\omega^{03}_{\phi}) \\ &&
0\eh d\theta \wedge d\phi \ (-r \cos(\theta)-\omega^{32}_{\phi} r) .\ea
At this point we make the following ansatz $ \omega^{\theta}_{01}=0,
\omega^{\phi}_{01}=0, \omega^{x}_{23}=0, \ {\rm and}\
\omega^{t}_{23}=0$.  From which we get the solution (\ref{resulti}).

\end{appendix}


\begin{thebibliography}{10}

\bibitem{observ}
  M.~J.~Reid,
  ``Is there a Supermassive Black Hole at the Center of the Milky Way?,''
  Int.\ J.\ Mod.\ Phys.\  D {\bf 18} (2009) 889
  [arXiv:0808.2624 [astro-ph]].
  %%CITATION = IMPAE,D18,889;%%
A.~Mueller,
  ``Experimental evidence of black holes,''
  PoS {\bf P2GC} (2006) 017
  [arXiv:astro-ph/0701228].
  %%CITATION = POSCI,P2GC,017;%%
 A.~E.~Broderick, A.~Loeb and R.~Narayan,
  ``The Event Horizon of Sagittarius A*,''
  Astrophys.\ J.\  {\bf 701} (2009) 1357
  [arXiv:0903.1105 [astro-ph.HE]].

\bibitem{wald}
  R.~M.~Wald,
  ``General Relativity,''
%\href{http://www.slac.stanford.edu/spires/find/hep/www?irn=1334239}{SPIRES entry}
{\it  Chicago, Usa: Univ. Pr. ( 1984) 491p}.


\bibitem{kerrnew}
R.~P.~ Kerr, 
``Gravitational field of a spinning mass as an example of algebraically special metrics''
Phys.\ Rev.\ Lett.\ {\bf 11} (1963) 237-238. E.~T.~Newman, R.~Couch, K.~Chinnapared, A.~Exton, A.~Prakash and R.~Torrence,
  ``Metric of a Rotating, Charged Mass,''
  J.\ Math.\ Phys.\  {\bf 6} (1965) 918.
  %%CITATION = JMAPA,6,918;%%

\bibitem{hawking}
  S.~W.~Hawking and G.~F.~R.~Ellis,
  %``The Large scale structure of space-time,''
%\href{http://www.slac.stanford.edu/spires/find/hep/www?irn=6991262}{SPIRES entry}
{\it  Cambridge University Press, Cambridge, 1973}

%\cite{beke}
\bibitem{beke}
  J.~D.~Bekenstein,
  ``Black holes and entropy,''
  Phys.\ Rev.\  D {\bf 7} (1973) 2333.
  %%CITATION = PHRVA,D7,2333;%%

\bibitem{Hawking:1974sw}
  S.~W.~Hawking,
  ``Particle Creation By Black Holes,''
  Commun.\ Math.\ Phys.\  {\bf 43} (1975) 199
  [Erratum-ibid.\  {\bf 46} (1976) 206].

\bibitem{string}
  A.~Strominger and C.~Vafa,
  ``Microscopic Origin of the Bekenstein-Hawking Entropy,''
  Phys.\ Lett.\  B {\bf 379} (1996) 99
  [arXiv:hep-th/9601029].

\bibitem{carlip}
  S.~Carlip,
  ``Entropy from conformal field theory at Killing horizons,''
  Class.\ Quant.\ Grav.\  {\bf 16} (1999) 3327
  [arXiv:gr-qc/9906126].
S.~Carlip,
  ``Black hole entropy from conformal field theory in any dimension,''
  Phys.\ Rev.\ Lett.\  {\bf 82} (1999) 2828
  [arXiv:hep-th/9812013].

\bibitem{bhe0} C.~Rovelli,
  ``Black hole entropy from loop quantum gravity,''
  Phys.\ Rev.\ Lett.\  {\bf 77} (1996) 3288
  [arXiv:gr-qc/9603063].

\bibitem{bhe1}
Ashtekar A, Baez B and Krasnov K.
``Quantum geometry of isolated horizons and black hole
entropy'' {Adv.\ Theor.\ Math.\ Phys.\ } {\bf 4} 1-94.


\bibitem{babaez}
  J.~C.~Baez,
  ``An introduction to spin foam models of BF theory and quantum gravity,''
  Lect.\ Notes Phys.\  {\bf 543} (2000) 25
  [arXiv:gr-qc/9905087].
  %%CITATION = LNPHA,543,25;%%


\bibitem{nous}
  J.~Engle, A.~Perez and K.~Noui,
  ``Black hole entropy and SU(2) Chern-Simons theory,''
  arXiv:0905.3168 [gr-qc].


\bibitem{kiril-lee}
K.~V.~Krasnov,
  ``On statistical mechanics of gravitational systems,''
  Gen.\ Rel.\ Grav.\  {\bf 30} (1998) 53
  [arXiv:gr-qc/9605047].
L.~Smolin,
  ``Linking topological quantum field theory and nonperturbative quantum
  gravity,''
  J.\ Math.\ Phys.\  {\bf 36} (1995) 6417.


%\cite{Ashtekar:1999wa}
\bibitem{ack}
  A.~Ashtekar, A.~Corichi and K.~Krasnov,
  ``Isolated horizons: The classical phase space,''
  Adv.\ Theor.\ Math.\ Phys.\  {\bf 3} (2000) 419
  [arXiv:gr-qc/9905089].
  %%CITATION = 00203,3,419;%%

  \bibitem{jon}
  A.~Ashtekar, J.~Engle and C.~Van Den Broeck,
  ``Quantum horizons and black hole entropy: Inclusion of distortion and
  rotation,''
  Class.\ Quant.\ Grav.\  {\bf 22} (2005) L27
  [arXiv:gr-qc/0412003].
  %%CITATION = CQGRD,22,L27;%%
  C. ~Beetle,  J.~ Engle "Quantization of generic isolated horizons in loop quantum gravity."
To appear.

\bibitem{better}
 A.~Ashtekar, C.~Beetle and J.~Lewandowski,
  ``Geometry of Generic Isolated Horizons,''
  Class.\ Quant.\ Grav.\  {\bf 19} (2002) 1195
  [arXiv:gr-qc/0111067]. 

\bibitem{ih_prl} Ashtekar A,  Beetle C, Dreyer O, Fairhurst S,
Krishnan B, Lewandowski J and Wi\'sniewski J 2000 Generic Isolated
Horizons and their applications \textit{Phys.\ Rev.\ Lett.}
\textbf{85} 3564-3567


\bibitem{lewa}
  J.~Lewandowski,
  ``Spacetimes Admitting Isolated Horizons,''
  Class.\ Quant.\ Grav.\  {\bf 17} (2000) L53.
  %[arXiv:gr-qc/9907058].
  %%CITATION = CQGRD,17,L53;%%


\bibitem{afk} Ashtekar  A,  Fairhurst  S and Krishnan  B 2000 Isolated
horizons: Hamiltonian evolution and the first law  \textit{Phys. \
Rev.} D \textbf{62} 104025

\bibitem{abl2002} Ashtekar  A, Beetle  C and Lewandowski J 2002
Geometry of generic isolated horizons  \textit{Class. Quantum Grav.}
\textbf{19}  1195-1225
%
% Among other things, this reference mentions that the isolated horizon
% boundary conditions generically select a unique equiv. class of null normals,
% and they also introduce the partial gauge-fixing of `good cuts'.
%

\bibitem{abl2001} Ashtekar  A, Beetle  C and Lewandowski  J
2001 Mechanics of rotating isolated horizons \textit{Phys.\ Rev.\ }
D \textbf{64}  044016

\bibitem{lew2000} Lewandowski J 2000
Space-times admitting isolated horizons
\textit{Class.Quant.Grav.} \textbf{17} L53-L59


\bibitem{thiemann}
T.~Thiemann,
  ``Modern canonical quantum general relativity,''
%\href{http://www.slac.stanford.edu/spires/find/hep/www?irn=7656084}{SPIRES entry}
{\it  Cambridge, UK: Cambridge Univ. Pr. (2007) 819 p}

\bibitem{cov} C.~Crnkovic and E.~Witten, in `Three hundred years of
  gravitation'; ed. S. Hawking, W. Israel. J.~Lee and R.~M.~Wald,
  ``Local symmetries and constraints,''
J.\ Math.\ Phys.\ {\bf 31}
  (1990) 725.
A.~Ashtekar, L.~Bombelli and O.~Reula,
 in '200 Years After
  Lagrange', Ed. by M. Francaviglia, D. Holm.

\bibitem{witten}
  E.~Witten,
   ``Quantum field theory and the Jones polynomial,''
  Commun.\ Math.\ Phys.\  {\bf 121} (1989) 351.
  %%CITATION = CMPHA,121,351;%%


\bibitem{tate}
 A.~Ashtekar,
  ``Lectures on nonperturbative canonical gravity,''
%\href{http://www.slac.stanford.edu/spires/find/hep/www?irn=2507340}{SPIRES entry}
{\it  Singapore, Singapore: World Scientific (1991) 334 p. (Advanced series in astrophysics and cosmology, 6)}

\bibitem{majundar}
 R.~K.~Kaul and P.~Majumdar,
  ``Quantum black hole entropy,''
  Phys.\ Lett.\  B {\bf 439} (1998) 267.
 R.~K.~Kaul and P.~Majumdar,
  ``Logarithmic correction to the Bekenstein-Hawking entropy,''
  Phys.\ Rev.\ Lett.\  {\bf 84} (2000) 5255.

\bibitem{models}
 E.~Livine and D.~Terno,
  ``Quantum black holes: Entropy and entanglement on the horizon,''
  Nucl.\ Phs.\  B {\bf 741} (2006) 131.

\bibitem{amit}
  A.~Ghosh and P.~Mitra,
  ``A bound on the log correction to the black hole area law,''
  Phys.\ Rev.\  D {\bf 71} (2005) 027502.
  %[arXiv:gr-qc/0401070].
  %%CITATION = PHRVA,D71,027
 G.~Gour,
  ``Algebraic approach to quantum black holes: Logarithmic corrections to black
  hole entropy,''
  Phys.\ Rev.\  D {\bf 66} (2002) 104022.
  
\bibitem{barba}
  I.~Agullo, J.~F.~Barbero G., J.~Diaz-Polo, E.~Fernandez-Borja and E.~J.~S.~Villasenor,
  ``Black hole state counting in LQG: A number theoretical approach,''
  Phys.\ Rev.\ Lett.\  {\bf 100} (2008) 211301.
  J.~F.~Barbero G. and E.~J.~S.~Villasenor,
  ``Generating functions for black hole entropy in Loop Quantum Gravity,''
  Phys.\ Rev.\  D {\bf 77} (2008) 121502.
  ``On the computation of black hole entropy in loop quantum gravity,''
  Class.\ Quant.\ Grav.\  {\bf 26} (2009) 035017
\bibitem{barberos}
  I.~Agullo, J.~F.~B.~G., E.~F.~Borja, J.~Diaz-Polo and E.~J.~S.~Villasenor,
  ``The combinatorics of the SU(2) black hole entropy in loop quantum
  gravity,''
  arXiv:0906.4529 [gr-qc].
  %%CITATION = ARXIV:0906.4529;%%


\bibitem{carlip-log}
  S.~Carlip,
  ``Logarithmic corrections to black hole entropy from the Cardy formula,''
  Class.\ Quant.\ Grav.\  {\bf 17} (2000) 4175
  [arXiv:gr-qc/0005017].


\bibitem{rezendeyo}
  D.~J.~Rezende and A.~Perez,
  %``4d Lorentzian Holst action with topological terms,''
  Phys.\ Rev.\  D {\bf 79} (2009) 064026
  [arXiv:0902.3416 [gr-qc]].


%\cite{Liu:2009em}
\bibitem{liu}
  L.~Liu, M.~Montesinos and A.~Perez,
  %``A topological limit of gravity admitting an SU(2) connection formulation,''
  arXiv:0906.4524 [gr-qc].
  %%CITATION = ARXIV:0906.4524;%%
%\cite{Ashtekar:1998ak}

\bibitem{noni} A. Perez, ``Black Hole
Entropy and SU(2) Chern  
Simons Theory'', International Loop quantum Gravity Seminar, 
http://relativity.phys.lsu.edu/ilqgs/perez050410.pdf
May 4, 2010. 

\bibitem{wi} A.~Corichi and E.~Wilson-Ewing,
  ``Surface terms, Asymptotics and Thermodynamics of the Holst Action,''
  arXiv:1005.3298 [gr-qc].



\bibitem{zapata}
  A.~Ashtekar, A.~Corichi and J.~A.~Zapata,
  ``Quantum theory of geometry. III: Non-commutativity of Riemannian
  structures,''
  Class.\ Quant.\ Grav.\  {\bf 15} (1998) 2955
  [arXiv:gr-qc/9806041].
  %%CITATION = CQGRD,15,2955;%%

\bibitem{lost}
J.~Lewandowski, A.~Okolow, H~Sahlmann, and Thiemann T.
\newblock Uniqueness of the diffeomorphism invariant state on the quantum
  holonomy-flux algebra.
\newblock 2004. C.~Fleischhack,
  ``Representations of the Weyl Algebra in Quantum Geometry,''
  Commun.\ Math.\ Phys.\  {\bf 285} (2009) 67
  [arXiv:math-ph/0407006].


\bibitem{chandra}
  S.~Chandrasekhar,
  ``The mathematical theory of black holes,''
%\href{http://www.slac.stanford.edu/spires/find/hep/www?irn=1845780}{SPIRES entry}
{\it  Oxford, UK: Clarendon (1992) 646 p.}

\bibitem{future1}  A. Perez, D. Pranzetti, in preparation.

\bibitem{beigin}  A. Perez, ``$SU(2)$ Chern-Simons theory and black hole entropy'', Plenary talk presented at LOOPS09, Beiging, China, August 2009.

\bibitem{future2} K. Noui, J. Engle, A. Perez, ''The SU(2) Black Hole entropy revisited'', in preparation.


%\cite{Kaul:2010kg}
\bibitem{maju}
  R.~K.~Kaul and P.~Majumdar,
  ``Schwarzschild horizon dynamics and SU(2) Chern-Simons theory,''
  arXiv:1004.5487 [gr-qc].
  %%CITATION = ARXIV:1004.5487;%%

\end{thebibliography}
\end{document}